\documentclass[twocolumn,apj,numberedappendix,twocolappendix]{OJAc}
\usepackage{newtxtext,newtxmath}

\usepackage[T1]{fontenc}

\DeclareRobustCommand{\VAN}[3]{#2}
\let\VANthebibliography\thebibliography
\def\thebibliography{\DeclareRobustCommand{\VAN}[3]{##3}\VANthebibliography}

\setcitestyle{notesep={; }, round, aysep={,}, yysep={;}, uniquename=init}
\usepackage{graphicx}	
\usepackage{amsmath}	
\usepackage{csquotes}
\usepackage{soul} 
\usepackage[spanish,es-minimal,english]{babel}

\usepackage{xcolor}

\begin{document}

\title{The past, present and future of observations of externally irradiated disks}

\author{Planet formation environments collaboration}\thanks{Corresponding author: Thomas Haworth, t.haworth@qmul.ac.uk}
\author{Megan Allen$^1$}
\author{Rossella Anania$^2$}
\author{Morten Andersen$^{3}$}
\author{Mari-Liis Aru$^{3}$}
\author{Giulia Ballabio$^4$}
\author{Nicholas P. Ballering$^{5,6}$}
\author{Giacomo Beccari$^{3}$}
\author{Olivier Bern\'e$^7$}
\author{Arjan Bik$^8$}
\author{Ryan D. Boyden$^6$}
\author{Gavin Coleman$^9$}
\author{Javiera Díaz-Berrios$^{10}$}
\author{Joseph W. Eatson$^1$}
\author{Jenny Frediani$^8$}
\author{Jan Forbrich$^{11}$}
\author{Katia Gkimisi$^{12}$}
\author{Javier R. Goicoechea$^{13}$}
\author{Saumya Gupta$^9$}
\author{Mario G. Guarcello$^{14}$}
\author{Thomas J. Haworth$^9$}
\author{William J. Henney$^{15}$}
\author{Andrea Isella$^{16}$}
\author{Dominika Itrich$^{17}$}
\author{Luke Keyte$^9$}
\author{Jinyoung Serena Kim$^{17}$}
\author{Michael Kuhn$^{11}$}
\author{Franck Le Petit$^{18}$}
\author{Lilian Luo$^{19}$}
\author{Carlo Manara$^{3}$}
\author{Karina Maucó$^{3}$}
\author{Rapha\"{e}l Meshaka$^{9,20}$}
\author{Samuel Millstone$^{16}$}
\author{James E. Owen$^4$}
\author{S\'ebastien Paine$^9$}
\author{Richard J. Parker$^1$}
\author{Tyger Peake$^9$}
\author{Megan Peatt$^{21}$}
\author{Paola Pinilla$^{19}$}
\author{Lin Qiao$^9$}
\author{Mar\'ia Claudia Ram\'irez-Tannus$^{22}$}
\author{Suzanne Ramsay$^{3}$}
\author{Megan Reiter$^{16}$}
\author{Ciarán Rogers$^{23}$}
\author{Giovanni Rosotti$^2$}
\author{Ilane Schroetter$^{7}$}
\author{Andrew Sellek$^{23}$}
\author{Leonardo Testi$^{12}$}
\author{Sierk van Terwisga$^{24}$}
\author{Silvia Vicente$^{25}$}
\author{Catherine Walsh$^{10}$}
\author{Andrew Winter$^{22}$}
\author{Nicholas J. Wright$^{26}$}
\author{Peter Zeidler$^{27}$}


\affiliation{Institutions are listed after the references.}



\begin{abstract}

Recent years have seen a surge of interest in the community studying the effect of ultraviolet radiation environment, predominantly set by OB stars, on protoplanetary disc evolution and planet formation. This is important because a significant fraction of planetary systems, potentially including our own, formed in close proximity to OB stars. This is a rapidly developing field, with a broad range of observations across many regions recently obtained or recently scheduled. In this paper, stimulated by a series of workshops on the topic, we take stock of the current and upcoming observations. We discuss how the community can build on this recent success with future observations to make progress in answering the big questions of the field, with the broad goal of disentangling how external photoevaporation contributes to shaping the observed (exo)planet population. Both existing and future instruments offer numerous opportunities to make progress towards this goal.
\end{abstract}




\section{Introduction}
This work follows a series of four meetings stimulated by multiple large grant awards related to the study of how radiation fields in stellar clusters affect protoplanetary disk evolution and planet formation in a process called `external photoevaporation'. These meetings were hosted as follows:

\begin{enumerate}
    \item University of Milan, Summer 2023, Giovanni Rosotti and the ERC StG DiscEvol. 
    \item ESO, Garching, Winter 2023, Carlo Manara and the  ERC StG WANDA.
    \item The Royal Society, London, Summer 2024, Thomas Haworth and the ERC CoG FRIED.
    \item The Rice Paris Institute, Paris, Winter 2024, Megan Reiter and NSF CAREER award.
\end{enumerate} 
This document aims to briefly summarise the state of the field as captured at those meetings, identify outstanding issues and propose some means via which progress can be made with existing and future facilities.  

\section{A brief review of the study of external photoevaporation}
\label{sec:review}

\subsection{Environmental influences on planet formation}

External photoevaporation is one of several processes through which the environment external to protoplanetary disks may influence the process of planet formation, with potential consequences for the resultant exoplanet population. 
Planet-forming disks are found around young stars (typically with ages $\lesssim 10$~Myr). This time scale for planet formation coincides with the period over which stars inhabit their nascent, hierarchically collapsing star forming regions \citep[e.g.][]{2003ARA&A..41...57L}. In such environments of enhanced local star and gas density, star-disk systems may be gravitationally perturbed by stellar encounters \citep[e.g.][]{2023EPJP..138...11C}, accrete from the ambient interstellar medium \citep[e.g.][]{2020NatAs...4.1158P} and, in the presence of nearby strong UV sources, lose material to thermal winds from the heated outer disc. The latter process is known as external photoevaporation \citep{2022EPJP..137.1132W}.

\subsection{Early studies}

\subsubsection{Discovery of proplyds}

\begin{figure*}
    \centering
    \includegraphics[width=2.0\columnwidth]{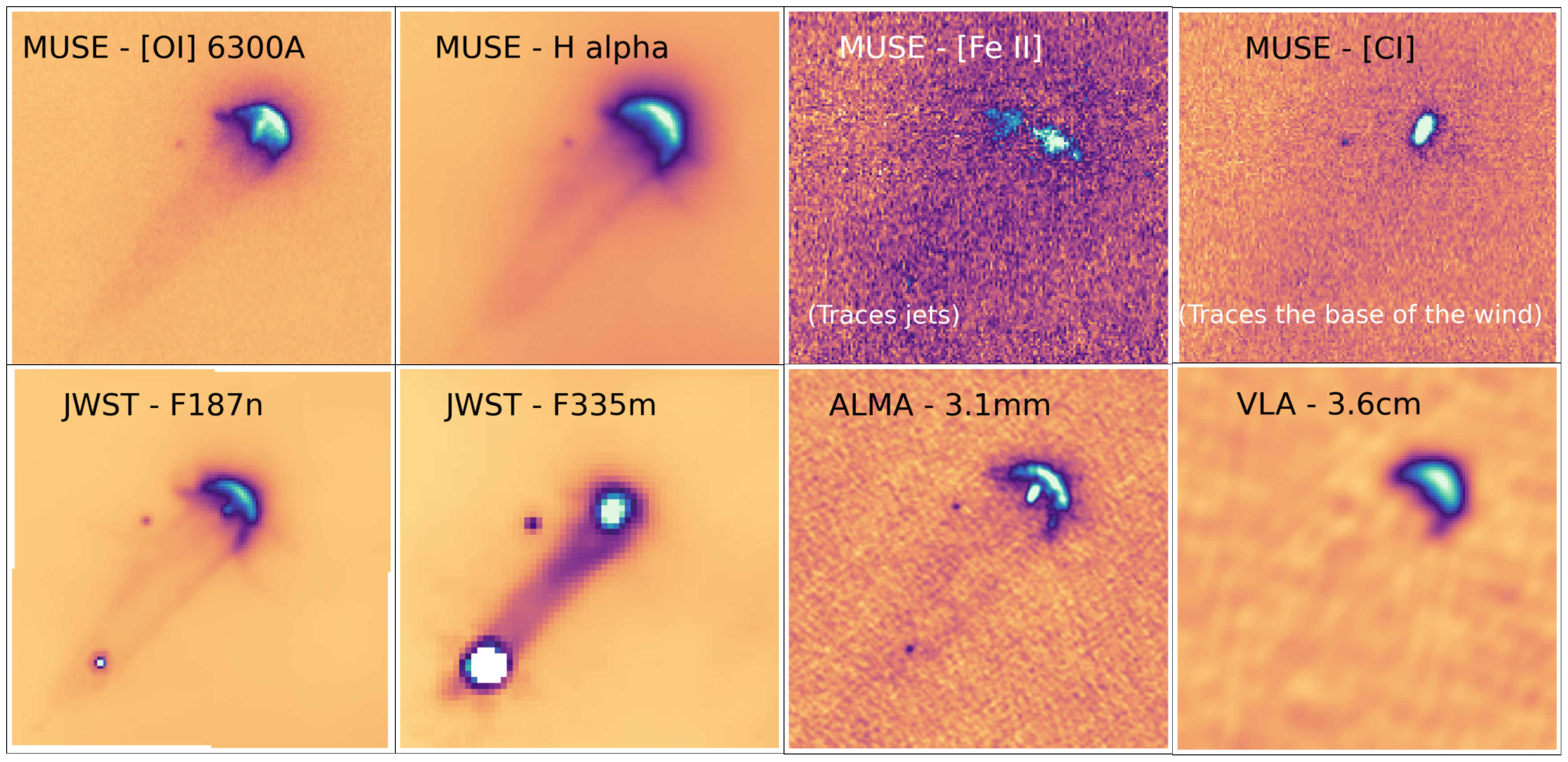}
    \caption{A gallery of images of the proplyd 177-341 in the ONC from different facilities. The field of view is $4\times4''$. The upper 4 panels are from VLT/MUSE \protect\citep{2024A&A...687A..93A}, which from left to right are the [O\,\textsc{i}]\,6300\AA\ line (tracing the disk, PDR and ionisation front), H\,$\alpha$ (predominantly tracing the ionisation front), [Fe\,\textsc{ii}] 8617\AA\ (tracing jets) and the [C\,\textsc{i}] 8727\AA\ which traces the base of the wind. The left two lower panels  are F187n and F335m from JWST NIRCam \protect\citep{2023arXiv231003552M}. The lower right panels are ALMA \protect\citep{2023ApJ...954..127B} (tracing dust in the disk and free-free emission from ionisation front) and VLA \protect\citep{2016ApJ...831..155S} (tracing the ionisation front) observations respectively. Note that ELT/HARMONI will offer around an order of magnitude higher spatial resolution over MUSE/NFM and JWST. SKA will also provide around an order of magnitude enhancement in spatial resolution compared to the VLA image here at similar wavelengths.   }
    \label{fig:gallery}
\end{figure*}

The first unambiguous evidence for the external photovaporation of protoplanetary disks came from Hubble Space Telescope (HST) observations of young stars in the Orion Nebular Cluster (ONC). While some stars showed clear disks in silhouette, many also had an associated teardrop morphology with a bright ``cusp'' pointed towards the UV sources in the region, which were the first identified external photoevaporative winds \citep{1993ApJ...410..696O, 1994ApJ...436..194O, 1998AJ....115..263O, 2000AJ....119.2919B}. The term \enquote{proplyd} was used to describe all PROtoPlanetarY Disks, but this terminology has evolved to now refer to the teardrop shaped externally irradiated disks/winds. Figure \ref{fig:gallery} shows a multi-wavelength, multi-instrument view of the ONC proplyd 177-341 over a $4\times4''$ field. It includes images from VLT/MUSE, JWST, ALMA and the VLA, which combined provide a deep insight into the properties of important different parts of the proplyd. We are now in an era where we have a comprehensive multi-wavelength view of many of these objects, particularly those in the relatively nearby ($\sim$400\,pc) Orion molecular cloud complex.

The early HST observations were followed by observations with facilities such as Keck \citep{1999AJ....118.2350H, 2000AJ....119.2919B} and stimulated theoretical work trying to understand the mass flux and observational characteristics \citep{1998ApJ...499..758J, Henney:1998-proplyd-brightness-profile-models, 1998ApJ...502L..71S, 2000ApJ...539..258R}. Theory and observations agreed that the rate of mass loss $\dot{M}_{\mathrm{wind}}$ from many proplyds is sufficiently high that the depletion time scale  $\tau_\mathrm{dep} = M_\mathrm{d}/\dot{M}_{\mathrm{wind}}$ for a typical disc mass $M_\mathrm{d}$ is very short ($\ll 10^5$
yr in some cases), giving rise to the ``proplyd lifetime problem'': why are objects that are predicted to be so short-lived observed in regions that are $>1$\,Myr in age \citep[e.g.][]{1999AJ....118.2350H}?

\subsubsection{Evolutionary context}

Not long after the discovery of proplyds, pioneering work interpreting photometry of young stars by \citet{1998AJ....116.1816H} revealed the fraction of stars with NIR excess or inner disk across the ONC decreased with distance from the centre of the cluster (decreasing external UV flux). While this trend is probably a consequence of the presence of younger stars towards the central regions \citep{1997AJ....113.1733H}, this suggested that external photoevaporation is not the dominant factor in dispersing this inner disc material in the ONC. Indeed $80$~percent of disks in the central $\sim 0.15$~pc survive, a fact that seemed to require that these disk-hosting stars have only recently migrated into the highly irradiated region \citep{1999ApJ...515..669S}. This turned out to be a solution difficult to reconcile with the nominal age of the ONC (typically estimated at $\sim 1-3$~Myr) unless discs remained very compact and/or implausibly massive \citep{2001MNRAS.325..449S}. On the other hand, compact disks would have much lower mass loss rates than apparent among the proplyds. 

Interest in the topic waned somewhat in the early 2000s, possibly reflecting the apparent inefficiency of external photoevaporation in influencing the inner disk material  probed by the photometric constraints of the time. 
From a theoretical perspective it was clear that external photoevaporation should preferentially drive mass loss in the outer disk, where the gravitational potential is shallowest, perhaps leaving the inner disk untouched. 
However, \citet{2004ApJ...611..360A} showed that winds can be launched even when the disk is entirely contained within the gravitational radius (the radius where gravitational potential balances thermal energy). Then \citet{2007MNRAS.376.1350C} applied the first viscous disk evolutionary models including external photoevaporation to show that the process truncates the disk down to the point where the wind loss and viscous spreading rates match, at which point the disk is depleted over a viscous time scale. Broadly, if the inner material is replenished from the outer disk, even the inner disk life time should at least somewhat depend on external UV field. 

Evidence for a gradient in disk life-times with increasing external UV field was only finally revealed when disk fractions across more distant star forming regions with several O stars were observed  \citep{2004AJ....128..765S,2012A&A...539A.119F, 2016arXiv160501773G, 2023ApJS..269...13G}. These studies  found fractionally fewer disk-hosting stars at high far ultraviolet (FUV) fields, indicating that an FUV flux $\gtrsim 10^3$\,G$_0$ could indeed shorten the inner disk lifetime. \citet{2008ApJ...675.1361F} estimated, based on the demographics of star forming regions, that the majority of stars (and planetary systems) form in such environments, suggesting that the external UV field influences the formation of the majority of the local exoplanet population. However, despite inducing mass-loss and shortening disk lifetimes, external photoevaporation does not necessarily suppress planet formation. \citet{Throop:2005-sedimentation-evaporation} suggested that preferential removal of gas from the outer disk may, in fact, be conducive to the growth of solid particles by concentrating dust. This idea was supported by evidence from the multi-wavelength opacity measurements of an edge-on silhouette disk in the ONC, 114-426 \citep{Throop:2001-grain-growth}. Nonetheless, the nature of the influence that external photoevaporation has on forming planets in different disk regions broadly remained (and remains) unclear.

\subsubsection{Anatomy of externally photoevaporating discs}

\label{sec:anatomy}

Equally as important as the evolutionary implications of external photoevaporation are the (micro)physics and observable signatures of the wind itself. Ground-based mid-infrared observations of the Orion Nebula 
\citep{Robberto:2005-mir-m42, Smith:2005-mir-m42}
revealed that the proplyd disks, ionization fronts, and wind-interaction arcs all emit brightly in the 10 to 30 micron range, with silicate dust features seen in emission 
\citep{Shuping:2006-mir-silicate-emission-proplyds}.
\cite{2008AJ....136.2136R} published a catalogue of all proplyds (and non-proplyd YSOs) across the ONC, giving a final number of 178 proplyds. 

At this time, a lot of work focused on proplyd models to diagnose the conditions and abundances in proplyds from optical line emission \citep{2005ApJ...621..328H, 2012MNRAS.426..614M, 2013MNRAS.430.3406T}.  \cite{2013ApJ...765L..38V} combined HST, VLT/NACO and VLT/VISIR observations to make the first estimate of polycyclic aromatic hydrocarbons (PAHs) abundance in the proplyd HST 10, finding that they appear to be less abundant than in the ISM. Finally, ahead of its time, \cite{2013ApJ...766L..23W} made the first consideration of the impact of external irradiation on disk chemistry, predicting that some line fluxes can be significantly enhanced by external irradiation, something that is now being studied, a decade later, with ALMA and JWST as we will discuss below.


Over the years, candidate proplyds were also suggested in different regions. Some of these were identified spectroscopically, for example, \citet{2009A&A...495L..13R} suggested the existence of external photoevaporation in $\sigma$ Ori based on VLT/FLAMES spectroscopy \citep[but these are still yet to be directly imaged/resolved, see also][]{2023A&A...679A..82M}. 
 {Others have been detected in the infrared, 
 such as the strongly illuminated proplyd \(\sigma\) Ori IRS~1B,
 initially discovered via mid-infrared silicate emission \citep{van-Loon:2003-sig-ori-irs-mir} 
 and confirmed via near-infrared H and He recombination line observations \citep{Hodapp:2009-sig-ori-irs-nir}}.
 
At longer wavelengths, an intriguing class of proplyd-like objects has been discovered in Spitzer/MIPS $24\,\mu\mathrm{m}$ images by \citet{Balog:2006-spitzer-mir-24-micron}, consisting of three dusty tails - in NGC 2244, IC 1396 and NGC 2264 - that seem to be severely gas-depleted \citep{Balog:2008-gas-free-proplyd-tails}, possibly representing a later stage of disk evolution.
So far, these seem to represent a rare class of proplyd with no more than 19 known candidates \citep{Balog:2006-spitzer-mir-24-micron,2008ApJ...687L..37K,2016ApJ...826L..15K,2019RNAAS...3...95T,2020RNAAS...4...15T}, mostly in more distant star-forming regions. The closest examples are KCFF-1 in NGC 1977 \citep{2016ApJ...826L..15K}, though this does show free-free emission from the ionisation front \citep{2024ApJ...967..103B} indicating gas is present, and two candidates surrounding LOri043 and LOri065 in the \(\lambda\) Orionis region \citep{2019RNAAS...3...95T}, which is the oldest region ($\sim6$ Myr) where proplyds have been suggested.

Finally, other objects suggested to be proplyds are often teardrop shaped objects on a much larger scale \citep{2000AJ....119..292B, 2003ApJ...587L.105S, 2012ApJ...746L..21W, 2012ApJ...761L..21S,2013ApJ...773..135G, 2014A&A...565A.107G, 2016ApJ...825L..16M} and are likely to be evaporating gaseous globules/globulettes (EGGs). Globules often contain young stellar objects (YSOs) and are very interesting in their own right \citep[see the above references and e.g.][]{2019MNRAS.490.2056R, 2020MNRAS.496..394R, 2020MNRAS.497.3351R,2020A&A...639A...1G}. However, YSOs embedded within globules are probably gaining mass from the globule, and so we typically distinguish them from proplyds, where the disk itself is losing mass by external photoevaporation.

From the mid 2010s everything changed quite dramatically in the study of planet formation and protoplanetary disks thanks to new observational facilities. Prior to this, the optical view of the silhouette disks in the ONC provided a small number of spatially resolved examples, which all showed an unstructured internal morphology \citep[e.g.][]{2000AJ....119.2919B, 2008AJ....136.2136R}. However, the unprecedented high resolution and sensitivity of ALMA at mm-wavelengths, as well as high contrast of ground-based near-IR imagers from 8-m telescopes, revealed that disks are, in fact, highly structured, with rings, gaps, asymmetries and other features in gas and dust \citep[e.g.][]{2015ApJ...808L...3A, 2018ApJ...869L..41A, 2021ApJS..257....1O, 2023ASPC..534..605B}. This has led to massive advances in our understanding of disk processes and planet formation, and continues to do so. However, the focus of high spatial resolution observations in this period was on nearby ($\sim130$--$190$\,pc) low mass and density star forming regions, such as Taurus/Lupus, where targets are brighter and easier to resolve. Large mm-sized dust grains probed by ALMA continuum observations are mostly well-settled into the disk mid-plane \citep[e.g.][]{2020A&A...642A.164V}, as well as being radially compact with respect to the gaseous disc \citep[e.g.][]{2017A&A...605A..16F}. In addition, even for molecular line emission, the high spatial resolution interferometry filters out large scale structure. These biases have made it easy to consider disks as isolated systems. Aside from systems with clear gravitational encounters (i.e. multiples), disk properties were not typically understood to be influenced by a dynamic external environment. 

\subsection{A new burst of activity}

From 2016 external photoevaporation experienced a surge of renewed interest. Given the large volume of work produced, we split progress over this period by theory and observation. We start with only a very brief summary of theoretical developments, given that this paper focuses on observations.  

\subsubsection{Theory}

Significant advances were made in understanding the important microphysics of an FUV driven wind, leading to substantial revision of the expected mass loss rates. \cite{2016MNRAS.457.3593F} calculated new semi-analytic solutions for far-ultraviolet radiation (FUV) driven mass loss, accounting for the fact that only small dust grains are entrained in the wind, which lowers the extinction to the disk outer edge and promotes mass loss. This was followed by the development of \textit{radiation hydrodynamic simulations} capable of modeling external photoevaporation \citep{2016MNRAS.463.3616H, 2019MNRAS.485.3895H} which was significant because they i) enabled more flexible exploration of the parameter space than semi-analytic models, ii) extended the dimensionality of the semi-analytic models, which were restricted to 1D and iii) enabled more realistic radiative transfer comparisons with observations (which we discuss in \ref{sec:newobs}).

An impactful outcome of the radiation hydrodynamic models was the public grids of mass loss rates that have made the study of external photoevaporation more accessible to the wider community \citep{2018MNRAS.481..452H, 2023MNRAS.526.4315H}. This in turn has led to many works studying disk evolution, including: exposure to constant UV fields \citep[e.g.][]{2018MNRAS.475.5460H, 2020MNRAS.492.1279S,
2022MNRAS.514.2315C, 2023A&A...674A.165W, 2024A&A...681A..84G, 2025MNRAS.539.1190C}, exposure to time varying UV fields in cluster calculations \citep[e.g.][]{2019MNRAS.490.5678C, 2019MNRAS.485.4893N, 2022MNRAS.512.3788Q, 2023MNRAS.520.5331W, 2021MNRAS.502.2665P}, and coupling of external photoevaporation to detailed models of planet formation and evolution  \citep[][]{2022MNRAS.515.4287W, 2022MNRAS.517.2103D, 2023MNRAS.522.1939Q, 2023A&A...673A..78E, 2024arXiv240719018H}. The earliest stages of planet formation also depend on the rate of drift of dust through the disk due to the `headwind' it experiences \citep{2012A&A...539A.148B}, and \citet{2020MNRAS.492.1279S} showed that the truncation of the disk by external photoevaporation results in rapid inward drift of mm-sized grains from the outer disk. Whether or not this dust survives depends on whether dust traps (i.e. pressure bumps in the disk mid-plane) can efficiently stop inward migration before the outer disk is depleted \citep{2024A&A...681A..84G}.

Regarding disk demographics in the solar neighbourhood, \cite{2018MNRAS.478.2700W} showed that external photoevaporation is important for disk evolution, and that dominates over random star-disk encounters in observed star forming regions. In the galactic context, other works suggested that external photoevaporation may dominate over internal disk dispersal mechanisms for a majority of disks \citep{2020MNRAS.491..903W,2020MNRAS.495L..86L}. 

\label{sec:newobs}

\subsubsection{Probing growing dust with mm-continuum interferometry}

Continuum surveys in the submillimetre (sub-mm) offer a powerful window into the demographic evolution of protoplanetary discs. If we can assume that the disk emission is optically thin, can be disentangled from free-free emission, and that the disk temperature is constant (usually $\sim 20$~K), then the flux from a given disk is a proxy for the total dust mass. In this way, SMA observations by 
{\cite{2010ApJ...725..430M}}
already showed evidence for depletion of disk dust masses in the highly irradiated core $\sim 0.3$~pc of the ONC compared to those at $\sim 1$~pc separations, despite the contrary influence of the UV heating on continuum flux \citep{2021MNRAS.503.4172H}. With ALMA, this became clearer for larger samples of disks in and around the ONC { \citep{2014ApJ...784...82M, 2018ApJ...860...77E, 2019A&A...628A..85V}}. A dust mass gradient as a function of distance from the main UV source was also observed in $\sigma$ Orionis by \cite{2017AJ....153..240A}, and median disk masses also appear to be anti-correlated with ambient UV broadly across the Orion complex down to very low FUV radiation fields \citep[G$_0$ of order unity][]{2022A&A...661A..53V, 2023A&A...673L...2V,2025A&A...693A..49S}. \cite{2020A&A...640A..27V} also used ALMA to identify a subset of disks which appeared to be depleted in the continuum, despite a high fraction ($\sim 80$~percent) of surviving disks based on NIR excess across the region
\citep{2001AJ....121.1512H}. This result appeared to echo the findings in the ONC, i.e. that cold millimeter dust is more easily depleted than inner disk material, and motivated follow up searches for proplyds (see Section~\ref{sec:wind_obs}).  

Information on the spatial distribution of the dust requires resolved continuum observations, which are more challenging for the irradiated disks in Orion than the comparatively nearby early ALMA targets. Even so, the high resolution surveys by \cite{2018ApJ...860...77E} and \citet{otter2021} demonstrated that mm continuum disk sizes in the core of the ONC are smaller than those in similarly aged low mass star forming regions. A possible interpretation for this finding is truncation or inward dust drift due to external photoevaporation \citep{2020MNRAS.492.1279S}. \cite{2024ApJ...976..132H} recently presented high resolution ALMA continuum observations towards disks in $\sigma$ Orionis. This study, along with some remarkable individual case studies, tentatively showed fewer substructures for the disks in higher UV parts of the cluster. However, such inferences are made extremely challenging by the inherent bias in high-resolution observations which tend to target sources with a substantial quantity of remaining dust.

\subsubsection{Tracing disc winds at optical and mm/cm wavelengths}
\label{sec:wind_obs}

\begin{figure*}
    \centering
    \includegraphics[width=0.9\linewidth]{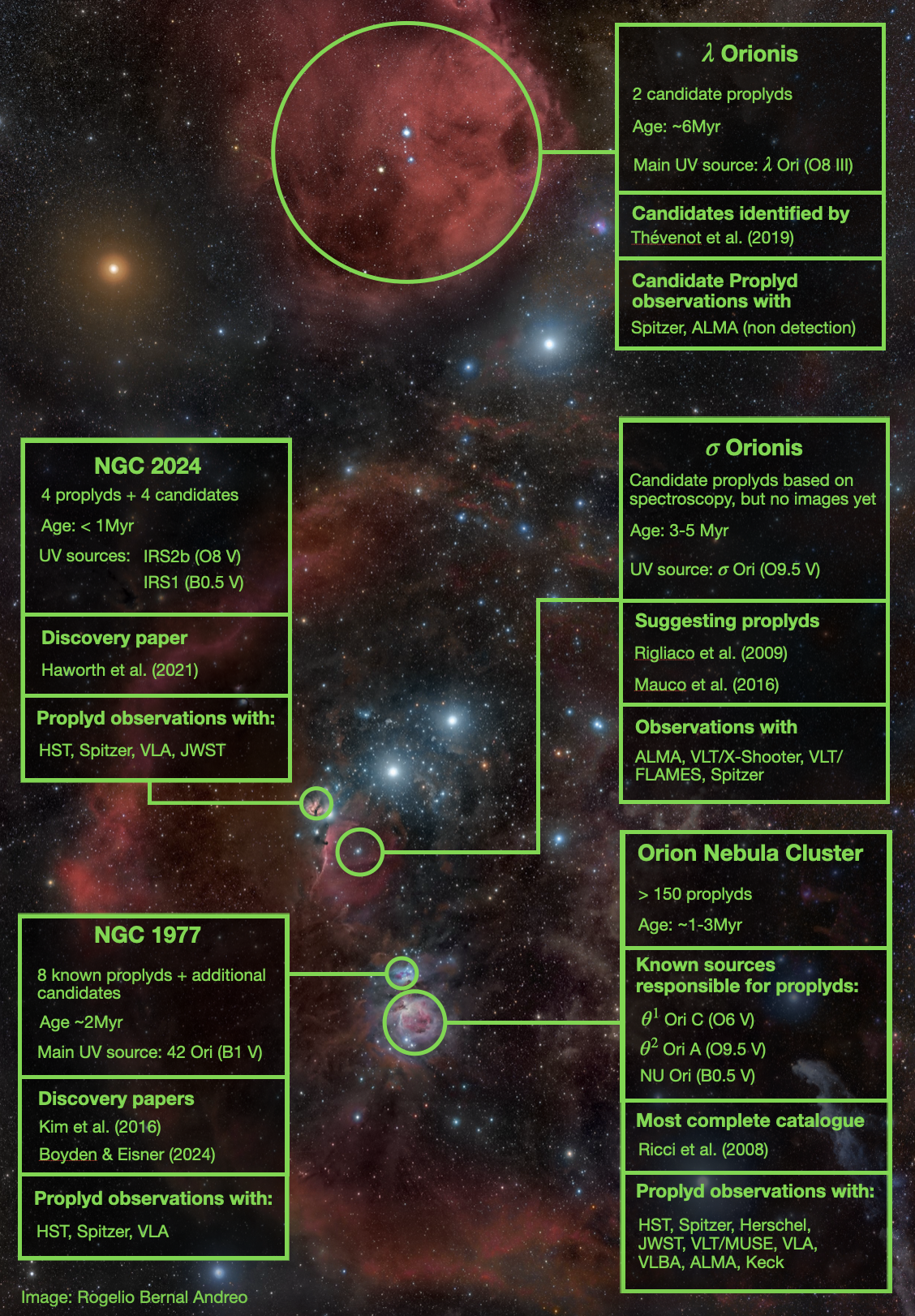}
    \caption{A summary of the regions across the Orion constellation where proplyds (or candidate proplyds) have been identified. All of these regions are at distances of around 400\,pc.}
    \label{fig:OrionProplyds}
\end{figure*}

 The presence and morphology of a proplyd ionisation front is probed either by highly ionised emission lines or free-free emission at radio wavelengths. Outside of the ONC, \cite{2016ApJ...826L..15K} discovered 7 proplyds in the vicinity of the B1V star system 42 Ori in NGC 1977 (with an age $\sim$1--2 Myr) using HST and Spitzer  \citep[note that one other proplyd, Parenago 2042, was discovered in NGC 1977 by][]{2012ApJ...756..137B}. The discovery of these eight proplyds was significant because it demonstrated that the reach of external photoevaporation extends to much weaker UV environments ($\sim10^3$\,G$_0$) compared with close proximity to the O6V star $\theta^1$~Ori~C in the ONC ($10^4-10^7$G$_0$). \cite{2021MNRAS.501.3502H} also discovered a handful of proplyds with HST in the very young region NGC 2024, demonstrating that disks can be irradiated at very young ages ($\sim$0.5--1 Myr), at a time that might even influence planet formation at early stages. \cite{2016ApJ...833L..16F} also discovered a planetary-mass proplyd in the ONC. Identifying proplyds in this way requires spatially resolved observations. Therefore, all of these proplyds are observed toward clusters at around 400\,pc in the constellation of Orion, as summarised in Figure \ref{fig:OrionProplyds}, rather than more distant regions. 

VLT/MUSE has also recently proven to be a powerful tool in the study of proplyds, and has the potential to be invaluable for identifying much more distant examples. As an integral field spectrograph, in the narrow field mode it provides low spectral resolution maps across the optical range at higher spatial resolution than offered by HST or the James Webb Space Telescope (JWST). At present, VLT/MUSE has been used to observe around a dozen proplyds \citep[][see Fig. \ref{fig:gallery} for an example]{2023A&A...673A.166K, 2023MNRAS.525.4129H, 2024A&A...687A..93A}, providing a unique view of their ionisation and photodissociation structures. These experiments are also valuable in the efforts to use line diagnostics to identify proplyds at distances at which they cannot be spatially resolved \citep[e.g. based on predictions such as those of][]{2023MNRAS.518.5563B}. However, these efforts are in the early stages \citep[with groundwork being laid by e.g.][]{2024A&A...685A.100I} and so we return to them later in this paper. 

At radio wavelengths, the VLA is proving a very useful tool in the identification and interpretation of proplyds. It enables us to search for signs of ionisation fronts over large fields of view, even when there is significant foreground extinction. \cite{2024ApJ...967..103B} presented a VLA survey towards NGC 1977, identifying many candidate new proplyds, though given the low ionising flux of 42 Ori follow up work is required to confirm that the signal is not due to internal ionisation by the host star (the star the disk orbits).   \cite{2023ApJ...954..127B} combined ALMA and VLA observations to separate the disk and ionization front contributions to the radio continuum (e.g. Fig. \ref{fig:gallery}). This meant they could use an emission measure to estimate the density in the ionisation front and hence provide a measure of the mass loss rate. This was in reasonable agreement (to within an order of magnitude) of the more commonly adopted approach of estimating the mass loss rate based on the assumption of photoionization equilibrium given the expected incident ionizing flux and ionization front radius \citep[see][]{2022EPJP..137.1132W}. There are additional VLA and VLBA surveys of the ONC with underutilised potential for the study of proplyds \citep{2016ApJ...822...93F, 2021MNRAS.506.3169V, 2021ApJ...906...23F}.

\subsubsection{Emission originating close to the disc surface}

Lower ionisation state and molecular gas emission lines typically probe the denser regions, often at the base of the wind or within the bound disc structure itself. Several bright lines of this kind fall within ALMA wavelengths, and are therefore detectable and in some cases spatially resolvable at the distance of Orion. \cite{2020ApJ...894...74B, 2023ApJ...947....7B} performed a HCO+ and CO line survey of disks in the core of the ONC (FUV fluxes $\gtrsim 10^4\, $G$_0$), finding that they are likely compact, massive and CO rich compared to disks in lower mass clusters. By contrast, at lower UV intensities ($\sim 500\,$G$_0$) \citet{2024ApJ...969..165D} infer similar molecular abundances to nearby low-mass SFRs. This hints that overall chemistry may only be influenced at the fairly extreme UV field strengths found in the central ONC.

Another interesting result of \citet{2023ApJ...947....7B}  is that the gas-to-continuum radius ratio appears smaller in the ONC compared to lower mass regions, suggestive of truncation \citep[e.g.][]{2024A&A...681A..84G}. Adding to the circumstantial evidence, in the broader Orion region \citet{2024A&A...685A..54V} find no resolved discs in their scattered light NIR observations with VLT/SPHERE in regions exposed to FUV field larger than 300\,G$_0$.

There has also been growing interest in probing neutral atomic carbon via its sub-mm fine-structure lines, which are a PDR tracer and predicted to be a good kinematic tracer near to the base of the wind \citep{2020MNRAS.492.5030H}. Yet, a search with APEX towards the NGC 1977 proplyds resulted in non detections only \citep[either due to low disk masses, carbon depletion or beam dilution,][]{2022MNRAS.512.2594H}. There have been a handful of detections of atomic carbon towards disks in low UV environments \citep[of order $1-10$G$_0$][]{2016A&A...588A.108K}, and there is now a single detection towards the 203-506 FUV irradiated disk near the Orion Bar \citep{2024A&A...689L...4G}. In the NIR, VLT/MUSE recently revealed that the forbidden line at [CI] 8727\AA\ is an excellent tracer of the  irradiated disk surface \citep{2023MNRAS.525.4129H,2024A&A...692A.137A}. {In general, NIR lines from electronically excited states of atomic carbon are excellent tracers
of external FUV  \citep{1991ApJ...375..630E,2024A&A...689L...4G}}. While the NIR lines cannot currently be spectrally resolved, they do offer potential diagnostic utility in finding externally irradiated disks when a proplyd morphology is not resolved.  

\subsubsection{Infrared constraints on PDR physics and inner disk survival}

There has been limited study of proplyds in the mid-to-far infrared.
\textit{Spitzer} has been valuable in the search for proplyds, but typically more so at the shorter wavelengths, where there is more archival data \citep[e.g. with IRAC,][]{2016ApJ...826L..15K}.
\cite{2017A&A...604A..69C} studied dense PDR lines with \textit{Herschel} towards four cometary objects, including two proplyds in the ONC. They studied FIR lines such as [OI]\,63\,$\mu$m and high-J CO lines to determine the conditions in the PDR of the flow, and independently estimate a mass loss rate for these proplyds that was consistent with prior estimates. \cite{2016ApJ...829...38M} also used \textit{Herschel} to study YSOs in $\sigma$ Ori, finding [NII] emission associated with one of the larger disks, but not for the smaller targets, possibly indicative of external photoevaporation.  We will discuss below how future FIR concept missions may offer new utility in the study of external photoevaporation.

A NIR excess is also an expected signature of a dusty inner disk, and is therefore used as a probe of disk dispersal. Many works have studied the inner disk fractions as a function of cluster mass and age, as well as a function of location (UV field) within the cluster \citep[e.g.][]{2004AJ....128..765S, 2011ApJS..194...10P, 2012A&A...539A.119F,
2015A&A...578A...4S, 2015ApJ...811...10R, 2016arXiv160501773G, 2018MNRAS.477.5191R, 2023ApJS..269...13G}. However these can be difficult to interpret for a number of reasons. For bulk disk fractions, comparisons between regions can be challenging because both age constraints and sensitivities (e.g. as a function of stellar mass) are often inhomogeneous. Even for disk fraction gradients inferred from single surveys within individual regions, care must be taken over the fraction of contaminants and age gradients \citep[e.g.][]{2014ApJ...787..109G}. While anti-correlations between disk fractions and FUV flux appear to be clear in at least some regions, this may not be causal, and could plausibly originate from an alternative (environmental) process.

\subsubsection{The terrestrial planet forming zone and PDR chemistry with JWST}

A more recent development is the introduction of JWST, providing high sensitivity and high resolution spectroscopy in the near- and mid-infrared. So far there have been two primary studies of externally irradiated disks with published results. One is the XUE project, which targeted 12 disks in NGC 6357 with MIRI/MRS. 
{Their sample contains disks around stars with spectral types fom late G to early F, making them borderline cases between intermediate-mass T-Tauri and Herbig disks \citep[][Ramírez-Tannus et al. in prep.]{2023SSRv..219....7B}}.
NGC 6357 is a more extreme region than the ONC, with multiple very massive stars, and containing one of the most massive stars in our Galaxy  \citep[Pis24-1, O4III(f+)+O3.5If*;][]{2003IAUS..212...13W}. They have so far published results for a first target \enquote{XUE~1} \citep{2023ApJ...958L..30R}{, a T-Tauri disk around a $\sim1$~M$_{\odot}$ star. They }find an inner disk bearing H$_2$O, CO, $^{12}$CO$_2$, HCN, and C$_2$H$_2$, in a very similar fashion to {T-Tauri} disks in nearby regions \citep[e.g.][]{2024PASP..136e4302H}. 
The rest of the XUE sample shows a very diverse chemical inventory, with spectra showing a rich molecular content, even for example in various isotopes of CO$_2$ (Frediani et al., \textit{in prep.}), while others seem to be depleted of water and other gas-phase molecules, in a fashion more similar to high-mass disks surveyed in nearby regions \citep{Fedele_2011, Banzatti_2020, Banzatti_2022}.
While the presence of PAHs is not homogeneous across the sample, all the spectra have a prominent 10~$\mu$m silicate feature, indicating that small, partially crystalline silicate dust is present at the disk surface (Ram\'irez-Tannus et al., \textit{in prep.}). This is similar to what is observed in samples of nearby T-Tauri stars, which is surprising given the very strong background radiation coming from the massive stars in NGC 6357. 
The authors hypothesize that the reason for not having enhanced line emission in the XUE sources is that these disks have been truncated due to external photoevaporation \citep{2025arXiv250400841P}.
{These results suggest} that perhaps the inner terrestrial planet forming zones are similar across a range of {external} UV environments. 

A second large study of irradiated discs is underway with the PDRs4All ERS collaboration, which obtained NIRCam, MIRI/Imaging, MIRI/MRS, and NIRSpec observations towards the Orion Bar and ambient parts of the ONC \citep{2022PASP..134e4301B}. Multiple papers have analysed the spectroscopic data on the intriguing FUV-only irradiated disk 203-506, revealing estimates of physical conditions and mass loss rate \citep{2024Sci...383..988B}, the first detection of the methyl cation CH$_{3}^+$, a key intermediate molecule in organic photochemistry \citep{2023Natur.621...56B}; the formation and destruction of water \citep{2024NatAs...8..577Z}; and the lack of volatile carbon depletion, along with a nearly solar C/O abundance ratio in the upper irradiated layers of the disk \citep{2024A&A...689L...4G}. 
The study of another proplyd in Orion (d203-504) with the PDRs4All data allowed the detection of water in the inner disk and the derivation of a gas-phase C/O abundance ratio of $\sim 0.5$ (close to Solar, Schroetter et al. submitted). A large  sample of proplyds is observed in photometry in a large number of narrow and broadband filters in the JWST survey of the inner Orion Nebula and Trapezium Cluster \citep[][see Fig. \ref{fig:gallery}]{2023arXiv231003552M, 2024arXiv241204356B} which is yet to be fully utilised.

From the above, we can see that the pace of work on external photoevaporation and the rate of new observations have accelerated dramatically in recent years, helped by the introduction of new facilities. We now discuss some of the key problems motivating this interest, before moving on to see what future observations are required to fully address those problems.

\section{Issues to address}
\label{sec:challenges}
The primary broader goal is of course to understand the complex interplay between all of the myriad processes that occur in protoplanetary disk formation and evolution, and with that to understand planet formation and exoplanet demographics. External photoevaporation of disks is just one part of that picture, but we have evidence that it can severely affect disk masses, radii and possibly lifetimes, at the very least in the case of the proplyds. So the main questions surrounding the impact of external photoevaporation in the context of the most aspirational goal mentioned above involve understanding what fraction of the total stellar population actually experience external photoevaporation, and whether they might do so at a time or in a manner that would influence planet formation. Here we explore specific goals or questions that need to be addressed to do with external photoevaporation. 

\subsection{What are the timescales and frequency of external irradiation?}
This is one of the most important outstanding problems in the study of external photoevaporation. We can look at a disk and see that it is subjected to a dramatic mass loss. We can run simulations where a disk is exposed to strong UV radiation and see that it predicts dramatic mass loss. However, from an observational perspective, what is not well understood is what fraction of stars are exposed to UV radiation fields of a given strength, and over what evolutionary period are they exposed to such a field? In particular, there is growing evidence and support for the idea that progress towards planet formation happens quickly \citep[e.g.][]{2018ApJ...857...18S, 2020Natur.586..228S, 2020A&A...640A..19T} so how do the timescales for irradiation and disk dispersal compare to possible planet formation timescales? This also ties in to the proplyd lifetime problem discussed above: why do we see proplyds when their depletion timescales are extremely short?

Addressing these questions really requires a connection to the processes of star formation and feedback. Stars form from collapsing clouds/cores, which in turn provide natal shielding of the star/disk from external irradiation. The point in time when any given star is stripped of that shielding depends either on the timescale for it to dynamically leave the surrounding cloud, or for the surrounding cloud to be dispersed by stellar feedback. Both of those timescales are non-trivial to determine, depending on the size of the ambient cloud, the local stellar number density (which may lead to ejection), the proximity of massive stars and overall timescale that feedback has been dispersing the molecular gas for in the region. 

There has been recent \textit{theoretical} work trying to address these questions by studying sink particles (stars) in star formation and feedback simulations. These in particular focus on the impact of ongoing star formation and natal shielding { \citep{2022MNRAS.512.3788Q, 2023MNRAS.520.5331W, 2023MNRAS.520.6159C, 2025MNRAS.536..298G}. 
}

The outstanding question is how to test this observationally? How can we quantify the impact of shielding? How can we determine when and to what extent disks were exposed to external irradiation in the past? There are at least two possibilities for making progress on this. One involves searching for very young externally irradiated disks, which reduces some of the complexity by stellar motions in the cluster but where we are limited in possible targets. NGC 2024 is one such very young region where this might be applied but others are much more distant, such as Tr 14 ($\sim2.7$\,kpc). In very young regions like these, determining the depleted and currently irradiated disk fractions directly places constraints on the timescale for irradiation. 

The more difficult alternative would require being able to interpret the dynamical history of the stars in a cluster and get a handle on the extinction over time. This requires astrometry and Gaia \citep{2016A&A...595A...1G} is enabling huge advances in this regard. However, a significant complication is that most massive stars are multiples which makes the overall proper motion vector of the multiple difficult to determine. A complementary approach here is to use N-body simulations to constrain the dynamical history of the cluster, for example as has been attempted in Cygnus OB2 \citep{2019MNRAS.485.1489W}, the ONC \citep{2019MNRAS.490.5478W} and NGC 2264 \citep{2022MNRAS.510.3178S,2022MNRAS.510.1136P}.

\subsection{How has external irradiation varied over cosmic time}
The peak star formation epoch was at \enquote{cosmic noon}, at a redshift of $\sim2$ \citep{2014ARA&A..52..415M}. Most of the exoplanet systems discovered in recent years {have ages between 2 -- 5\,Gyr \citep{2016A&A...585A...5B,2024AJ....167..270S} and therefore}  will have formed in environments more representative of star formation at such an epoch {\citep{2012MNRAS.426.3008K}}, rather than the present day, nearby Galactic regions that are often the focus of demographic studies of planet formation. 

Two important factors may have resulted in a different kind of planet formation at $z\sim 2$. 
One is that since the star formation rate was higher, probably the interstellar UV field was also stronger \citep{2014ARA&A..52..415M, 2020ARA&A..58..661F, 2020MNRAS.491..903W, 2024arXiv240809319H}. Another is that metallicity was lower. There is some evidence that disk fractions are observed to be lower at lower metallicity \citep[e.g.][]{2024ApJ...970...88P}. The reason for this is not fully understood, though it may be that internal photoevaporation is more effective at lower metallicity \citep{2010MNRAS.402.2735E, 2019MNRAS.490.5596W}. Stellar feedback also disperses molecular gas more rapidly at lower metallicity \citep[e.g.][]{2021MNRAS.501.4136A}, which may decrease the length of time for which disks are shielded by the ISM. Mass loss from protoplanetary disks due to external photoevaporation may also depend on metallicity. However, while this possibility has not been well studied, preliminary models suggest any such dependence is not a strong one \citep{2023MNRAS.526.4315H}. 

Making progress in understanding this observationally requires improving our understanding of disk evolution in relatively nearby regions of low metallicity.  For example \cite{2024ApJ...977..214D} find \textit{longer} lived disks in the low metallicity cluster NGC 346 in the Small Magellanic Cloud. More nearby, there is also a collection of lower metallicity star forming regions in the outer Galaxy such as Dolidze 25 \citep[e.g.][]{2021A&A...650A.157G, 2023JApA...44...67A}. However, a major obstacle is that even at these distances (e.g. to Dolidze 25 is 4-6\,kpc) we cannot resolve the disks. We therefore must interpret how disk  evolution proceeds based on spatially unresolved line emission. This unresolved line emission can be confused with background nebulosity in massive star forming regions. This underlines the importance of (and challenges in) developing new observational techniques for probing unresolved disk populations \citep[e.g.][]{2024A&A...684L...8R, 2024A&A...685A.100I}. 


\subsection{The interplay between infall and stellar feedback/external photoevaporation}
\label{sec:infall}

This question is not strictly one of external photoevaporation of disks, however streamers and infall formed a regular part of the discussion by the final meeting that stimulated this work. Infall is relevant to understanding the process of external photoevaporation observationally, because it may result in a similar environmental dependence of disk properties.

Infall has gained attention in recent years due to the discovery of large scale structures that appear to be signatures of infall onto disks \citep{2020NatAs...4.1158P, 2023A&A...670L...8G}. These accretion flows onto the disk are referred to as
\enquote{streamers}. In some cases, it has been determined that the kinematics of the gas is consistent with infall \citep{2024A&A...683A.133G} and that an observable shock is induced at the interception between disk and streamer \citep{2022A&A...658A.104G}. While infall is expected for very young stars ($\lesssim 0.5{-}1$~Myr) as part of the star formation process, there have been several protoplanetary disks with large scale structures that are possible signs of infall \citep[][]{2023A&A...670L...8G, 2023ApJ...943..107H, 2024MNRAS.528.6581H}. If they are infalling, the short free-fall time-scale for these $\sim 1000$~au structures ($\lesssim 1$~percent of the disk lifetime) implies such events are either much more numerous than observed, or continuously replenished (or probably some mix of both). These observations have led to the idea that \enquote{late stage} infall may be a significant driver of disk evolution \citep{2023EPJP..138..272K, 2024ApJ...972L...9W}. This idea is also circumstantially supported by prevalent misalignment between inner and outer disks \citep{2024ApJ...961...95V} and correlations between local ISM density and accretion rate \citep{2024A&A...691A.169W, 2024arXiv241205650R}, with the familiar caveats regarding possible age gradients. 

A key point is that streamers/candidate streamers have only been detected in low mass star forming regions, and are yet to be detected in more extreme environments. Demographics of the disks in Taurus, Chamaeleon, and Orion star-forming regions from SPHERE \citep{2024A&A...685A..53G, 2024A&A...685A..52G, 2024A&A...685A..54V} in fact reveal a relatively large fraction of objects in evident interaction with the environment for Taurus and Chamaeleon ($\sim15\%$), whereas no similar features are seen in Orion. {It is unclear if this is genuine or an observational bias, however there are some physical grounds to expect a possible difference.} In clusters bearing massive stars, feedback from winds, radiation and eventually supernovae creates hot low density bubbles and disperses the molecular gas 
\citep[e.g.,][]{2019Natur.565..618P,2020PASP..132j4301S,2020A&A...639A...2P}. Massive star feedback hence not only directly operates on disks, but could reduce the extent and the duration of late stage infall for many systems. To the authors knowledge, this issue is yet to be studied in the context of disk evolution.




\subsection{How does external irradiation affect disk composition}
\label{sec:disk_chemistry}
The disk composition is important both because it can provide information about the conditions and processes within disks (e.g. through chemical probes of the density/temperature and through processes like ice transport on dust), but also because planets will accrete atmospheres from disk gas and so atmospheric abundances may be sensitive to the disk chemistry. 

ALMA has provided a major shift in our ability to study the composition of the outer regions of disks. There is a wide array of studies but the impact is most concisely reflected in the success of ALMA large programs. The MAPS large program provided a high spatial and spectral resolution view of five disks in over 50 lines from over 20 different molecular species \citep{2021ApJS..257....1O}. The composition of these disks was very diverse and could also be affected by structure within the disk \citep[e.g.][]{mapsIII_law_2021}. The ongoing large program DECO (PI: Cleeves) is gaining a broader view surveying the chemistry (in particular determining the C/O ratio) of 80 disks across four regions.  Another large program, CHEER, that will provide a chemical study of Herbig disks was also awarded in Cycle 11 (PI: Pegues). AGEPRO (PI: Zhang) will also provide a large sample of disk mass and radii for different ages. 

More recently, JWST has enabled us to probe the composition of the inner regions of isolated disks. For example, the MINDS \citep{2024PASP..136e4302H} GTO program and JDISCS collaboration \citep{2024ApJ...963..158P} have surveyed large numbers of disks, and DECO has a JWST counterpart to study connections between the inner and outer disk chemistry, for example due to transport of species as ices by radial drift. Protostars, where the disks are still embedded in their collapsing envelopes and thus may be shielded from external radiation, have also been extensively surveyed by e.g. the JOYS GTO program, which found little evidence for molecular emission analogous to protoplanetary disks suggesting that their molecular spectra are dominated by the inner "hot core" of the envelope \citep{2024A&A...692A.197V}. {A catalog of proplyd spectra has also recently been obtained with JWST NIRSpec as part of cycle 3 \citep{rogers2024spectral} (PI: Rogers), providing $1-5\; \mu m$ spectroscopy for $\sim30\%$ of the known proplyds in the ONC. These observations will provide accretion and mass loss rates for a large sample of externally irradiated disks, as well as directly probe inner disk chemistry, and how it may be influenced by the external UV environment.}
{For cycle 4, a medium ($\approx 114$ hours) JWST program has been approved to observe 77 proplyds in Orion with MIRI/MRS (PI: I. Schroetter, co-PI: R. Boyden, O. Berné) and will provide the complementary $5-28~\mu$m wavelength coverage, allowing chemical diagnostics over a representative sample of externally UV-irradiated disks. This will provide a direct measurement of the gas kinetic temperature via the pure rotational lines of molecular hydrogen and will also allow for the detection of UV-processed chemistry such as OH, CH$_3^+$ and PAHs. Dust loss and chemistry will also be studied in detail in MIRI/MRS observations for three proplyds in the ONC (PI: N. Ballering).  }

The idea that external irradiation might affect the composition is certainly not new. \cite{2013ApJ...766L..23W} demonstrated that external irradiation can substantially alter the chemical composition of the outer disk, producing changes in both molecular abundances and integrated line intensities. These line intensities typically differ by factors of a few compared to isolated systems - a result of both modified chemical abundances and enhanced molecular excitation in the warmer gas. Similar effects are seen in models of disks exposed to moderate (1--100\,G$_0$) external radiation, where species such as C, C$^+$, CS, and N$_2$H$^+$ show significant changes in the outer disk \citep{gross2025}. However, despite the limited number of self-consistent chemical modeling studies exploring externally irradiated disks, there appears to be general agreement that the inner regions can remain effectively shielded from external radiation, maintaining compositions similar to isolated systems. While the exact extent of this shielded region is model-dependent, the qualitative picture is consistent, with column densities that vary by only factors of a few compared to isolated disks inside some radius \citep[see also][]{ndugu2024}.

The observational picture is more complex and requires careful interpretation. Recent work by \cite{KeyteHaworth2025} shows that photoevaporative winds can dominate line emission compared to UV-driven chemical effects, making it crucial to account for wind contributions when analyzing marginally-resolved sources. Observational studies have found varying degrees of external irradiation effects: \cite{2023ApJ...947....7B} found reduced CO depletion in ONC disks compared to those in lower UV environments, while \cite{2024ApJ...969..165D} and \cite{2023ApJ...958L..30R} reported relatively normal compositions for disks in the ONC outskirts and the inner disk of {XUE~1 in NGC 6357}, respectively. In contrast, the d203-506 and d203-504 disks in the ONC exhibit distinctive photochemistry and gas processing, as evidenced by CH${_3^+}$, near-IR CI, and PAH emission \citep{2023Natur.621...56B, 2024Sci...383..988B, 2024A&A...689L...4G}. There hence appears to be a diversity in how much external irradiation can affect the chemistry, and it is also unclear which parts of the disks are (or are not) affected. 

Overall, while theoretical models predict that external UV radiation should affect disk composition, observational evidence presents a mixed picture - some systems show clear signatures of photochemistry while others appear largely unaffected. Understanding this diversity will require both a larger sample of observations spanning different UV environments and careful selection of molecular diagnostics. Future JWST and ALMA surveys can address this by targeting specific tracers: line ratios such as CN/HCN and [CI]/N2H+ are sensitive to external FUV fields \citep{2013ApJ...766L..23W, gross2025}, while emission from [OI], [CI], OH, and high-$J$ CO lines primarily trace photoevaporative winds \citep{KeyteHaworth2025}. This combined approach - broader environmental sampling and selective observational tracers - will help disentangle the effects of external irradiation from wind emission and clarify how external irradiation shapes disk chemistry.

\subsection{The fate of dust in external photoevaporative winds}
The entrainment of dust in external photoevaporative winds is an important part of the process both i) because it limits the solid mass reservoir for planet formation and ii) because the dust entrained in the wind acts as an opacity source to the incident UV and can regulate the mass loss rate. 

Calculations of entrainment have been carried out in 1D
\citep{Throop:2005-sedimentation-evaporation, 2016MNRAS.457.3593F, 2021MNRAS.508.2493O, 2024A&A...681A..84G}, 
finding that only small grains are entrained in the wind and that there is also expected to be a radial gradient in maximum grain size near the disc outer edge. Dust trajectories in 2D simulations of external photoevaporation have now also been computed by \cite{2025MNRAS.539.1414P} finding similar behaviour near the mid-plane, but with even lower dust-to-gas ratios and maximum entrained grain sizes above the mid-plane, even in the scenario where all grain sizes could make it to the base of the wind. 

These theoretical studies are consistent with the limited observations we currently have on the dust in the outer disk/inner wind. \cite{2012ApJ...757...78M} studied dust in the outer regions of the 114-426 silhouette disk in the ONC, finding evidence for a gradient in grain sizes. Other silhouette disks in the ONC appear to show streaks of higher extinction emanating from the disk outer edge, but not from the disk surface, as expected from \cite{2025MNRAS.539.1414P}.  {Analysis of the mid-infrared emission from wind-interaction shells associated with proplyds in the inner Orion Nebula \citep{Garcia-Arredondo:2001-proplyd-bow-shocks} suggests that the grain opacity per hydrogen atom in the proplyd photoevaporation flow is depleted by as much as a factor of 30 compared with typical ISM values. However, this is based on observations at a single wavelength and is sensitive to the assumed grain temperature. Detailed modeling of recent JWST observations \citep{2023arXiv231003552M} should allow much tighter constraints on dust in winds to be established.}

Dusty proplyds \citep{Balog:2006-spitzer-mir-24-micron,2008ApJ...687L..37K,2016ApJ...826L..15K,2019RNAAS...3...95T} are also evidence that dust can be efficiently be removed from discs. These have often been regarded as more likely to be more evolved, gas-poor debris disks that are being eroded, however, the presence of an ionization front traced by free-free emission for KCFF-1 \citep{2024ApJ...967..103B}, suggests that there are some objects with significant mass loss in both dust and gas.

There are still many outstanding issues in the study of this entrained dust. For example, as it moves outward through the wind the density, temperature and UV all change. That may affect the grain structure, perhaps causing fragmentation that would further influence the attenuation of the UV. Desorbed gas species in the wind may also provide useful diagnostics of the grain composition in the outer disk. 

Another key component is Polycyclic Aromatic Hydrocarbons (PAHs), which are ubiquitous in the interstellar medium and play a central role in gas heating \citep{bakes1994photoelectric, berne2022contribution} and resulting mass-loss rates \citep{2023MNRAS.526.4315H}. Rare infrared spectroscopic observations (e.g. with VLT/VISIR, JWST) have confirmed the presence of PAHs in two Orion proplyds \citep{vicente2013polycyclic, shuping2014curious, 2024Sci...383..988B}. However, the extent of their distribution, their precise abundance, and whether their presence varies with the proplyds' age or the ambient radiation field remain uncertain.

\subsection{What is the impact on planet formation and exoplanetary system architectures?}

We mentioned above that some consideration is now being given to the possible impact of external photoevaporation on the resulting planets, however, this comes predominantly from a theoretical perspective \citep{2022MNRAS.515.4287W, 2023MNRAS.522.1939Q, 2024arXiv240719018H}, and there is not yet a clear prediction for a smoking gun signature of external photoevaporation in the planetary architectures. Really the main question that it boils down to is whether the planet formation and evolutionary timescales compare to the timescales on which the outer disk is destroyed by external photoevaporation, as well as the nature of the connection between the inner and outer disk. 

\cite{2020Natur.586..528W} showed that exoplanet properties, and particularly the fraction of hot Jupiters, depends on the density of stars in galactocentric position-velocity space. This could be the result of an environmental influence or age correlations, since hot Jupiters may inspiral onto the host star over time \citep[e.g.][]{2023AJ....166..209M}. It is less clear how an age correlation could explain correlations between occurrence rates of other classes of planets and galactocentric kinematics \citep[e.g.][]{2021AJ....162...46D}. Recently \citet{2023AJ....165..262Z} also showed that metallicity differences are insufficient to explain these variations. An environmental influence on planet formation is one possible explanation. For example, \citet{2024arXiv240809319H} suggest that lower planet fractions at higher galactocentric orbital amplitude may be the result of harsher UV environments at cosmic noon, which was when thick disk stars formed.

Convincing observational evidence of a causal relationship between environment and planet properties requires connecting planet-hosting stars to their birth environment. One possibility is to study young ($<100$\,Myr) exoplanetary systems. Transit and radial velocity surveys recover fewer planets around young stars both because such stars are comparatively rare and more active, while comparatively large uncertainties in stellar properties also translate to greater uncertainties in planet properties \citep[e.g.][]{2007ApJS..173..143B, 2019ApJ...880L..17N, 2024MNRAS.531.4275B, 2024AJ....167..210V}. Nevertheless, they may provide a means to study planetary architectures at a time where their cluster membership/possible environment is easier to identify. Furthermore, their planets will have been subject to less dynamical processing (e.g. due to planet-planet scattering) and atmospheric photoevaporation by the central star \citep{2017ApJ...847...29O} therefore differences between planets in low and high mass clusters may be more easily identified. 

Another aspect of this question is whether we can learn more through deeper study of disks in the process of undergoing external photoevaporation. For example, inner disk composition may be altered by external photoevaporation, as can now be probed by JWST (Section~\ref{sec:disk_chemistry}), and could then leave signatures observed exoplanet atmospheres. This kind of chemical footprint could also influence the abundance of elements in primitive bodies of the solar system. Alternatively, dust substructures in disks are often suggested to be associated with planet formation in disks in the low mass clusters \citep[e.g.][]{2018ApJ...869L..47Z}. Therefore, comparing substructures in disks as a function of radiation environment may therefore provide another empirical way to study the impact of external radiation field on planet formation. This is challenging but possible in Orion star forming regions with ALMA \citep{2024ApJ...976..132H}.

\subsection{Down to how weak a UV radiation field is external photoevaporation effective?}
Spatially resolved proplyds offer unambiguous evidence of external photoevaporation. However, in order to understand the true extent of external photoevaporation we also need to understand if and how it operates in lower UV radiation environments, where there may not be an ionisation front. There is evidence from the SODA survey \citep{2023A&A...673L...2V}  that the median dust disk mass is continuously increasing as a function of FUV environment even down to $\sim1$G$_0$. This result is surprising given that we may expect internal dispersal processes (e.g. planet formation, accretion, winds) to dominate the disk evolution at low external UV fields. It remains an important goal to identify the externally driven winds in such systems to verify that the explanation is (or is not) external photoevaporation. CO is not considered to be the best direct tracer of the wind itself because it is in a component of the wind where the kinematics only deviates marginally from that of the Keplerian disk. Atomic carbon has been suggested as an alternative which probes further out in the wind \citep{2020MNRAS.492.5030H}, and the strength of its emission lines seem correlated with external irradiation \citep{2024A&A...689L...4G,2024A&A...692A.137A}. Searching for winds with spatially and spectrally resolved atomic carbon observations with ALMA will provide a further test. Other diagnostics/identifiers of external winds in low UV environments should be developed. The recent work of \cite{2025arXiv250118752A} will alo be valuable here, which provides a new and more accurate way of quantifying the FUV radiation field in stellar clusters, and provides a catalogue for nearby clusters. 


\subsection{What is the explanation for anomalously old disks?}
In the context of external irradiation there are two main classes of \enquote{anomalously old} disks. One is the well known issue mentioned above that proplyds have short depletion timescales ($<<1$\,Myr) based on current mass loss rate estimates yet are observed in regions that are $\sim1-3$\,Myr in age (or perhaps even older if the candidate evaporating disks in $\sigma$ Ori or the dusty proplyd candidates in $\lambda$ Ori are real). This \enquote{proplyd lifetime problem} now has plausible resolution through one or more episodes of ongoing star formation, stellar dynamics in clusters, and the fact that the mass loss rate is a non-linear function of disk outer radius \citep[e.g.][]{2019MNRAS.490.5478W}. So the proplyd lifetime problem is no longer really much of a problem. For example, \cite{2022A&A...662A..74G} shows a specific example of a disk with a candidate embedded planet, recently emerging into a high UV environment at around $5\,$Myr. This is another example where the kinematics of cluster members will be valuable for obtaining a deeper understanding of the history and future of external irradiation in any given region. 

What is much less certain is the explanation for extremely old disks in lower UV environments -- the so-called Peter Pan disks \citep[e.g.][]{2018MNRAS.476.3290M, 2019ApJ...872...92F, 2020ApJ...890..106S}. These show signs of accretion even at ages of many tens of Myr. Recently there was also a suggestion of $\sim16$\,Myr old disk which is so flared and massive that it appears similar to very young disks \citep{2024arXiv240604160D}. \cite{2020MNRAS.496L.111C} demonstrated that the existence of Peter Pan disks could be explained so long as they are exposed to a weak external UV field, as well as ineffective internally driven winds, which tends to favour their survival around lower-mass stars \citep{2022MNRAS.509...44W}. However, Chandra observations from \cite{2022ApJ...935..111L} suggest that Peter Pan disks have neither an exceptionally low internal X-ray luminosity nor an exceptionally weak (current) external radiation field. Instead, late stage infall has long been suggested as a solution to these long-lived disks \citep{2014A&A...566L...3S}, and appears to an increasingly plausible scenario (as discussed in Section~\ref{sec:infall}). However, in this case relatively low degrees of angular momentum transport, internal wind mass loss and external UV irradiation are presumably also an important requirement. 
The first JWST spectrum of such a disk also showed its inner disk to have an particularly high C/O$\sim3$ \citep{2025ApJ...978L..30L}  indicating very carbon-rich gas relative to Solar and the ISM.

\subsection{Quantifying the uncertainties on measured parameters of externally photoevaporating disks}
The most obvious example of a key empirical parameter for externally photoevaporating disks that has poorly constrained uncertainties is the mass loss rate. In particular, when based on the ionisation front radius \citep[see][]{2022EPJP..137.1132W} the estimate assumes photoionisation equilibrium, uses the projected separation on the sky of the proplyd from the UV source and assumes no extinction. None of those uncertainties or assumptions are well quantified. Again, a clear improvement here will come from improved understanding of the 3D positions of stars in clusters. An alternative is to find multiple ways of estimating the mass loss rate and compare them. \cite{2023ApJ...954..127B} did so for a small sample of proplyds in the ONC using both the ionisation front radius and by using the radio emission measure to get the density at the ionisation front, finding that the two approaches were in good agreement (within an order of magnitude for most cases). Importantly, both mass loss rate estimates for proplyds from \cite{2023ApJ...954..127B} would indicate that external photoevaporation significantly affects the disk evolution. Additional estimates of the mass loss rates by other means with different assumptions will be valuable here to provide a measure of uncertainty, for example through radio recombination lines {\cite{2025arXiv250302979B}}.

\subsection{Was the pre-solar nebula an externally irradiated disk?}
Among the most commonly asked of questions when talks on external photoevaporation are given is \enquote{can you say anything about the birth environment of the Solar system}. To our knowledge most of the discussion on this topic centers on abundances of short-lived radionuclides (e.g. \citealt{gounelle2008origin}), and whether this requires proximity to and enrichment by a massive star, perhaps suggesting a strong UV environment was likely. However, there is a tension between the timescales required for enrichment from a massive star(s) and the timescales for the destruction of disks due to external radiation. \citet{2023MNRAS.525.2399P} show that significant destruction of disks may occur before enrichment to Solar System levels. \cite{2010ARA&A..48...47A} provides a detailed review on the topic of birth environment of the Solar system. More recently, \cite{2024arXiv240910638D} also provides a review from the perspective of meteoritics, and \citet{bergin2024interstellar} also discuss the question of the environment from several perspectives in connection with the formation history of the Solar System.


Given that we do not yet have unambiguous predictions from models for imprints of external photoevaporation on planetary architectures, it is difficult to say much from the perspective of planet formation models. There are well tailored models describing the solar system formation, so the better approach is probably to ask the question of how those would be affected if an external radiation field of given strength were introduced. There may be chemical imprints introduced in that case, for example (Okamoto et al. in prep) suggest that the anomalously high abundance of noble gases in the Jovian atmosphere might be related to external photoevaporation. Ultimately, though, we (or at least this subset of the community) cannot say with much certainty at this stage which constraints external photoevaporation puts on the formation environment of the Solar system. Progress here might be driven by new collaboration/discussion with teams specialising in Solar system formation.

\subsection{A comparative study of proplyds}
This is less of an outstanding question and more highlighting a significant gap in the literature. We now know of hundreds of proplyds, but only a handful have mass loss rate estimates. It would be beneficial to compare proplyds across regions in a consistent manner \cite[e.g. similar to the approach of][]{2024A&A...687A..93A}. To calculate their mass loss rates as a function of irradiation/disk properties and further to compare against and test the expectations of theoretical estimates of models such as the \textsc{fried} grids \citep{2023MNRAS.526.4315H}.

\subsection{Determining the conditions in proplyds}
Direct measures of the density, temperature and gas kinematics in different parts of a proplyd place valuable constraints on numerical models. Furthermore, estimates of these parameters and knowledge of the surface geometry can yield an estimate of the mass loss rate. 

Optical line ratios are used to determine the electron density and temperature in H\,\textsc{ii} regions \citep{2006agna.book.....O}. Optical line modeling of some proplyds was undertaken by e.g. \cite{1999AJ....118.2350H, 2013MNRAS.430.3406T}. However, the density in the ionised parts of proplyds is sufficiently high that collisional de-excitation limits their applicability, at least in commonly adopted line ratios. It is to be determined whether there are alternative diagnostics that could be employed to determine the conditions in the ionisation front. There has been recent success in constraining the conditions in the PDR using the \textsc{meudon} code \citep{2024Sci...383..988B}, but typically with a single representative value of density. 
{\cite{2025arXiv250302979B}}
also recently detected radio recombination lines for proplyds in the ONC, with which they can estimate the density and temperature near the ionisation front. There is much more work to do to in finding diagnostics that probe from the wind base through to the ionisation front. \\

\subsection{The connection to globules}
Globules have already been mentioned in our brief review in Section \ref{sec:review}. In high UV environments these often have a cometary proplyd-like morphology. However we make the slight distinction from \enquote{proplyds} because while YSOs embedded in a globule are still possibly gaining mass, externally photoevaporating disks are losing it. Yet, since the globule itself may disperse, leaving behind the disk, there must be some connection between the two. For example this may help explain the unusually large ONC proplyd 244-440 \citep[e.g.][]{2023A&A...673A.166K}. Furthermore, mass loss of the globule itself limits the mass reservoir for the YSO/disk to draw upon, which may affect both the properties of the central star(s) as well as their disks. At present, little work has been done on the connection between the two, and future multi-wavelength studies of proplyds and their embedded YSOs to determine their future evolution \citep[e.g. as][have done with VLT/MUSE, ALMA and VLT/ERIS]{2019MNRAS.490.2056R, 2020MNRAS.496..394R, 2020MNRAS.497.3351R, 2024MNRAS.527.3220R} would be very valuable. This also fits in with the discussion on late stage infall (Section~\ref{sec:infall}), since a disk embedded in and co-moving with a globule may continue to accrete in addition to being shielded from external photoevaporation.












\section{Comments on future observations: with and beyond the current state of the art}
\label{sec:futureobs}
Here we briefly discuss some highlights of what the community believes is required in the future to make significant progress in addressing some of the key issues raised in section \ref{sec:challenges}. This will not be exhaustive and the suggestions here are not written to be at a level suitable for a proposal. Rather, we hope to motivate the work among the community to flesh these ideas out into competitive proposals.

\subsection{Advances with current facilities}

\medskip

\subsubsection{Identifying and quantifying ongoing external photoevaporation in distant clusters}
\label{sec:IDdistant}
One of the main issues raised in Section \ref{sec:challenges} is the need to develop a picture of external photoevaporation in more massive clusters at a range of ages and metallicities. This is necessary to understand how widespread is the impact of external photoevaporation. However, these clusters are more distant and external photoevaporative winds cannot be spatially resolved within them.

To address this, emission line diagnostics are being developed to identify ongoing external photoevaporation in distant clusters. These are developed both from theoretical models \citep[e.g.][]{2023MNRAS.518.5563B} and by comparing the emission line properties of proplyds with disks in low UV environments \citep[e.g.][]{2024A&A...687A..93A}.  These could be searched for in optical line surveys of YSOs in distant clusters, for example using VLT/MUSE. This is being attempted towards Tr 14 \citep{2024A&A...685A.100I} and that work is really pioneering the major task of dealing with the nebula contamination that is present throughout clusters bearing massive stars. That will help facilitate future more extensive surveys (e.g. the large VLT/MUSE survey PID 112.25LK, PI Reiter) in massive clusters. 

The above optical line surveys would allow us to identify ongoing external photoevaporation by identifying excesses in optical lines. This might enable the calculation of the instantaneous fraction of disks subject to external photoevaporation. 

Another valuable approach to studying external photoevaporation in distant, high UV, clusters would be surveys of disk continuum mass estimates with ALMA. This would enable a comparison of disks throughout the different environments of a given cluster, but also between clusters (including those  nearby and already subject to study). This would also be very valuable for targets such as the XUE sample in NGC 6357 \citep{Ramirez-Tannus_2023}, where the interpretation of JWST observations implies compact disks and for which ALMA continuum mass estimates would confirm that interpretation.\\

\noindent\fbox{
    \parbox{\columnwidth}{
      \textbf{Suggested observations: External photoevaporation in distant clusters \\}
      \begin{itemize}
          \item Optical/IR spectroscopic surveys of YSOs in distant (out to $\sim2.5$\,kpc) clusters with the techniques to isolate the YSO/nebular emission and identify externally photoevaporating disks. [e.g. VLT/MUSE, NIRSpec] \\
          \item ALMA surveys of distant  (out to $\sim2.5$\,kpc) clusters to determine disk continuum mass distributions and compare with nearby low UV environments. \\
      \end{itemize}
    }
}

\subsubsection{Searching for additional proplyds}
Finding additional resolved proplyds is not a stamp collecting exercise. Each new cluster where they are identified tells us more about the prevalence of external photoevaporation, the timescales it happens upon, and how it may differ depending on the UV sources and environment. In particular there is great value in obtaining a complete census in individual regions since that places limits on the instantaneous irradiated disk fraction. 

There are still resolvable proplyds out there waiting to be discovered with existing facilities. NGC 1977 and NGC 2024 are regions with recently discovered proplyds \citep{2016ApJ...826L..15K, 2021MNRAS.501.3502H} but only a small fraction of those regions has actually been surveyed. We hence require larger scale surveys at wavelengths sensitive to external photoevaporation (e.g. to ionisation fronts). We also require observations that are able to identify proplyds in regions of high extinction (as in the inner part of NGC 2024 for example). There is also a complete dearth of high resolution irradiated gas tracer observations towards $\sigma$ Orionis and $\lambda$ Orionis, which may both host resolvable proplyds.

radio observations, for example with the VLA, provide a way of searching for ionisation fronts over large areas and at high extinction. New candidates were identified in this manner in NGC 1977 by \cite{2024ApJ...967..103B}. However, follow up observations are required to confirm their proplyd nature (e.g. the signal could be due to irradiation by the disk central star in those cases given that the B star 42 Ori is not a particularly strong emitter of extreme ultraviolet, EUV, photons).

Ideally we  want complete surveys of NGC 2024, NGC 1977, $\sigma$ Ori and $\lambda$ Ori in ionisation-front tracing lines at high spatial resolution, e.g. with VLT/MUSE, HST, or JWST (e.g. with NIRCam in the F187N filter, tracing Paschen $\alpha$.).

A key example for further study is the young region NGC 2024, which is so young ($<1\,$Myr) that any proplyds are likely being dispersed on a timescale that might influence planet formation \citep[e.g.][]{2018ApJ...857...18S,2023MNRAS.522.1939Q}. Knowing the evaporating (or evaporated) disk fraction at that early time therefore enables us to gauge how widespread the impact on the resulting planet populations might be. However, at present we have only patchy observations in paschen $\alpha$ and the foreground extinction is sufficiently high that the optical is inappropriate for a wider survey. JWST observations (PI: Meyer) of the very central region appear to reveal new candidate proplyds, but the field of view is only a small fraction of the cluster and there are no F187N observations in that dataset. \\

\noindent\fbox{
    \parbox{\columnwidth}{
            \textbf{Suggested observations: Finding new proplyds \\} 
      \begin{itemize}
          \item Radio surveys to identify candidate proplyds across NGC 2024, NGC 1977, $\sigma$ Ori, $\lambda$ Ori [e.g. VLA, VLBI] \\
          \item Complete optical and/or infrared imaging surveys of ionisation-front tracing lines at high spatial resolution to confirm and constrain mass 
          loss rates for new proplyds [e.g. HST, VLT/MUSE,Gemini/GMOS, JWST, ROMAN] \\
      \end{itemize}
    }
}

\subsubsection{Improving our understanding of known proplyds}
Figure \ref{fig:gallery} shows just part of the multi-wavelength view that we can now obtain of proplyds, probing from the disk, through the various stages of the wind. This provides more than one way of estimating the mass loss rate \citep{2023ApJ...954..127B} and with a large enough sample of objects covered in similar detail would enable us to directly test theoretical estimates of the mass loss rate as a function of UV environment, disk mass/radius and stellar mass \citep[e.g. from][]{2023MNRAS.526.4315H}. 

Radio recombination lines have recently been demonstrated by 
{\cite{2025arXiv250302979B}}
to provide estimates of the density/temperature near the I-front. This new approach should be applied to a wider range of systems to determine how diverse the conditions are close to the ionisation-fronts of proplyds. 

The use of JWST-NIRSpec in the multi-object spectroscopy (MOS) mode offers unique opportunities to extend the near infrared diagnostics discussed above to larger samples (hundreds) of disks. This can provide access to key tracers of the physics and chemistry for a representative sample of objects.

Proplyds have also been under-studied at very high spectral resolution. For example CRIRES+ and UVES with $R\sim100,000$ should be able to kinematically resolve the wind down to a few km\,s$^{-1}$. Multiple slits across key proplyds in different UV environments would provide new insight into the physical and velocity structure of the winds. This also provides key information on where and how the material is initially driven from the disk. The effective spatial and spectral resolution and large number of lines could enable the internal and external photoevaporative winds to be distinguished observationally.   \\

\noindent\fbox{
    \parbox{\columnwidth}{
            \textbf{Suggested observations: Improving understanding of known proplyds} 
    \begin{itemize}
          \item Consistent estimates of disk radii (e.g. ALMA), mass loss rates from (optical/IR, radio continuum) and density/temperature (e.g. from radio recombination lines) for a large sample of proplyds to provide the tightest constraints on models of external photoevaporation. 
          \item High resolution CRIRES+/UVES spectroscopy with multiple slits over a sample of key proplyds to build a map of the kinematic structure of proplyd winds down to $\sim$few\,km\,s$^{-1}$ velocities. \\
      \end{itemize}
    }
}

\subsubsection{Studying externally irradiated disk chemistry}
As discussed above, ALMA and JWST have both now been demonstrated as key tools for the study of chemistry of externally irradiated disks. However, there is the mystery that while some targets show clear evidence of photochemistry and chemistry affected by radiation environment \citep{2023ApJ...947....7B, 2023Natur.621...56B, 2024Sci...383..988B,2024NatAs...8..577Z, 2024A&A...689L...4G}, others do not and appear \enquote{normal} \citep[e.g.][]{2023ApJ...958L..30R, 2024ApJ...969..165D}. 

To understand why there is an apparent diversity in how the external radiation field affects disk chemistry we require a survey targeting a large number of star/disk/radiation field parameters with JWST and ALMA well beyond the handful of objects studied thus far. This combination will also enable us to determine whether there is a connection between the inner/outer disk chemistry (e.g. through radial drift) and whether that is also sensitive to environmental radiation field.  

While this is possible with existing ALMA and JWST as proved by the studies mentioned above, externally irradiated disks are more distant and hence require higher sensitivity observations, and are not so easily resolved as for disks in lower UV environments that have been studied in chemical surveys like the MAPS/DECO large programs. The future ALMA Wideband Sensitivity Upgrade (WSU) will massively improve our ability to survey the outer disk chemistry of disks in higher UV environments.

{These efforts on the observational side should be complemented by further developments in modeling predictions of promising signatures of environmental effects on disk chemistry, when and where they occur, and what the connection between the inner and outer disk should be. We identify potential future focuses among the community as including 2D and 3D thermochemical modeling of disk photoevaporative winds, the consideration of non-equilibrium chemistry in these flows, investigations of isotopic fractionation effects, and the impact of a time-dependent model for external irradiation. A key aim in the near future is also to quantify the impact of an altered outer disk, in terms of its physical, thermal, and chemical structure, on the transport of molecular species to the inner disk, and determine the dependencies on e.g. the properties of the central host stars.}

\noindent\fbox{
    \parbox{\columnwidth}{
            \textbf{Suggested observations: Chemistry of externally irradiated disks} 
      \begin{itemize}
          \item A large ALMA+JWST survey of chemistry in externally irradiated disks to determine in what scenarios the inner/outer disk chemistry is affected by radiation environment.
          \item The planned ALMA wideband sensitivity upgrade will enable larger scale surveys of disk chemistry in a range of UV environments. 
      \end{itemize}
    }
}
\subsubsection{Irradiation histories and cluster kinematics}
A key issue is understanding how the UV field distribution varies over time. In principle {\it Gaia} could help with this for clusters with parallax and proper motion measurements. However, it is challenging to use {\it Gaia} in massive clusters. They are typically quite extincted, however {\it GaiaNIR}  may help with this (see section \ref{sec:GaiaNIR}). Another challenge is that massive stars are often in multiples, which can affect the inferred proper motions of the UV sources.   We hence require a combination of Gaia analysis with detailed dedicated observations of the OB stars at high resolution to constrain their orbit parameters and proper motions (or at least the impact on the Gaia proper motion vector). For example this is being done for the 42 Ori multiple in NGC 1977 (PI: Kim) using Magellan MagAO-X imager and VIS-X spectrograph \citep{Haffert-2024}. 

The point of determining these irradiation histories is to understand what fraction of discs are irradiated, and when. There are two possible alternatives to achieving similar science outcomes. One would be to target very young clusters where the dynamical history is less complicated. The other may be to target runaway OB stars, where the motion of the UV source dominates over the other cluster dynamics and is predicted to leave clearer spatially varying imprints of external photoevaporation upon the disc properties (Coleman et al. submitted, where 42 Ori is studied as a possible example). \\

\noindent\fbox{
    \parbox{\columnwidth}{
        \textbf{Suggested observations: Irradiation history/cluster kinematics}       \begin{itemize}
          \item Detailed astrometry of massive star multiplicity and dynamics to isolate that from the overall proper motion of the multiple that may not be accounted for with Gaia. For example, with MagAO-X AO imaging and spectroscopy. \\
          \item Search for signatures of external photoevaporation near runaway/walkaway OB stars using high resolution imaging. 
      \end{itemize}
    }
}

\subsubsection{Proplyd accretion rates}
Another key aspect of disk evolution is accretion. One of the key general problems in protoplanetary disk studies is understanding the processes that facilitate accretion, particularly the debate on viscous vs MHD wind driven angular momentum transport. Obtaining simultaneous accretion measures and measures of the winds could help determine the relative impact of different processes on disk evolution. 

A simultaneous strength and weakness of proplyds is that the mass loss rates we measure for them are those from all sources. That is an internal photoevaporative or MHD is counted alongside an external wind when estimating the mass loss rate based on the radius of the ionisation front. Proplyd external photoevaporative mass loss rates combined with accretion rates may therefore provide a new way to determine how disks are evolving. The proplyd mass loss rates are straightforward to calculate, however accretion rates are more difficult. JWST NIRSpec-MIRI spectra can be used to do this based on the NIR luminosity (Schroetter et al. submitted) or combinations of Paschen $\alpha$ and Bracket $\beta$ following \cite{2024A&A...684L...8R}. \\

\noindent\fbox{
    \parbox{\columnwidth}{
        \textbf{Suggested observations: Accretion rates}       \begin{itemize}
          \item NIRSpec/MIRI spectra to provide accretion rates for all proplyds with resolved ionisation fronts ($\sim50$ targets across the ONC, NGC 1977, NGC 2024) \\
      \end{itemize}
    }
}

\subsubsection{Dust in external photoevaporative winds}
As discussed above, theoretical models can predict what sizes of dust grain are entrained in external photoevaporative winds \citep{2016MNRAS.457.3593F, 2021MNRAS.508.2493O} including in multidimensional radiation hydrodynamic simulations \citep{2025MNRAS.539.1414P}. Those latter models can also be used to predict what density, temperature and radiation field the grains will reside in as they propagate outwards in the wind. Those can be used to determine how the dust composition evolves in the wind. 

We can observationally test these predictions with combined JWST and HST observations of the outer disk, e.g. as in the case of 114-426 both using HST extinction to constrain grain sizes \cite{2012ApJ...757...78M} and JWST emission to study gas/ice composition \cite{2024arXiv241204356B}. Dusty tails have also been imaged with Spitzer/MIPS so JWST/MIRI is the natural successor to follow up these studies at higher resolution and at multiple wavelengths. \\

\noindent\fbox{
    \parbox{\columnwidth}{
        \textbf{Suggested observations: Dust in winds}       \begin{itemize}
          \item HST and JWST observations of proplyds in silhouette to study dust in the outer disk/inner wind in absorption \\
          \item JWST spectroscopy to study dust/ices in absorption/emission for proplyds. \\
      \end{itemize}
    }
}

\subsubsection{The interplay between feedback and infall}
Streamers/late infall are subject to increasing attention in recent years. If they are important processes, as is being suggested in low mass star forming regions, they will have to compete with stellar feedback in massive clusters. While not directly related to external photoevaporation we feel it sufficiently important to note that searches for infall/streamers at the periphery of H\,\textsc{ii} regions will be important to determine whether late infall/streamers are predominantly restricted to low mass star forming regions. \\

\noindent\fbox{
    \parbox{\columnwidth}{
        \textbf{Suggested observations: Late infall/streamers}   
      \begin{itemize}
          \item Measures of the fraction of embedded YSOs as a function of cluster mass and cluster age. 
          \item Intermediate baseline observations of YSOs that would reveal large scale structures for a statistically significant sample of YSOs in Orion, to compare with those already detected towards Taurus, Lupus, Rho Oph and so on.
          \item Searching for correlations between stellar accretion rate and ISM properties in irradiated regions. 
      \end{itemize}
    }
}

\subsection{Potential value of future concepts}
We now briefly summarise some upcoming facilities and what they may bring to the study of external photoevaporation.

\subsubsection{VLT/blueMUSE and VLT/MAVIS}

VLT/MUSE has demonstrated its value for the study of proplyds, offering integral field spectroscopy over 4650-9300\AA. Two new instruments for VLT will build on this scientific capability.

VLT/blueMUSE is a concept in development that would offer similar capability over the range 3500-5800\AA\ \citep{2019arXiv190601657R}. blueMUSE would provide information on key lines from He, C, N, O, Si and Mg that will provide tighter constraints on the physical conditions and abundances in proplyd winds.

The [OII] doublet at 3726\AA\ and 3728\AA\ can be used to trace the density in the ionization front, tracing densities between $\sim$10 - 10$^4$ cm$^{-3}$. In more extreme UV environments blueMUSE will also provide probes of high ionisation states. The parameters and abundances of massive stars and their ambient ISM is already identified as a key science case for blueMUSE \citep{2019arXiv190601657R} and the same approach will also yield new insight into proplyds. A key caveat of blueMUSE relative to MUSE is that it is not expected to have adaptive optics capability and the spatial resolution will be around a factor 2 worse than MUSE Narrow Field Mode. However, it will have twice the spectral resolution of MUSE on average and will have also twice the field of view (2 arcmin$^2$), allowing for simultaneous observations of many more protoplanetary disks in the same sky area. Its higher sensitivity, coupled with the better spectral resolution, will benefit the detection of emission lines within those objects. 

VLT/MAVIS will extend the existing adaptive optics capabilities of the VLT at visible wavelengths offering both imaging over a 30”x30” field of view and an integral field spectroscopic mode with a field of view of 5” x 7.2” with 40x50mas spaxels -- so a similar FOV and around a factor 2 smaller spaxel size than VLT/MUSE. 
 
 
\subsubsection{Extremely Large Telescope Instrumentation}

The instruments offered on the thirty-metre class Extremely Large Telescopes \citep[ELTs,][]{2022JATIS...8b1510S} will offer a combination of angular resolution and sensitivity that will be complementary to, and even exceed, that obtained with JWST and ALMA. The ELTs will provide extremely high spatial resolution observations of externally photoevaporating disks and will also provide very high spectral resolving power to complement that available with JWST at infrared wavelengths. The ELTs will enable us to resolve the disks for proplyds in distant clusters, which are at the limit of what ALMA can currently achieve, given the combined resolution/sensitivity requirements. ELTs may hence provide our best opportunity to compare disk radii in massive clusters like Carina with nearby star forming regions. 

With first light planned before the end of the decade, the initial instrument complement of ESO's ELT will provide adaptive optics supported imaging and spectroscopy from 1.0\,$\mu$m to 13\,$\mu$m with MICADO \citep{2024arXiv240816396S} and ELT/METIS \citep{2024SPIE13096E..12B}. MICADO offers a field of view of 50"x50" with 4mas pixels (18"x18" with 1.5mas pixels). The imaging modes of METIS provide a 10.5"x10.5" field from 3-5\,$\mu$m and 13.5"x13.5" from 8-13\,$\mu$m with angular resolution at the diffraction limit of the telescope ($\sim$20mas--70mas over the 3-13\,$\mu$m range). Not only will this enable the sharpest view of well studied proplyds in Orion (about 1.5\,au resolution) but more crucially will enable us to spatially resolve them in distant clusters. For example, we will get an ONC proplyd-like view out to the distance of the Carina star forming complex and be able to survey these more distant regions using a few telescope pointings. MICADO is expected to reach astrometric accuracy of at least 50mas, complementing GAIA for faint, embedded sources. 

This will enable us to go beyond spatially unresolved line fluxes to try and determine whether external photoevaporation is happening in distant clusters. Diagnostic spectral lines will be observable using slit spectroscopy in MICADO and METIS (spectral resolving power from a few 1000 to 20,000). Integral field spectroscopy with $R\sim 100,000$ is offered by METIS in the L and M bands with a field of view of just under 1". The workhorse spectrograph, HARMONI \citep{2024SPIE13096E..14T} also offers integral field spectroscopy with fields of view well matched to the characterisation of individual proplyds. The smallest field of view matches that of METIS at just under 1"-square with spatial sampling of 4mas. The largest field of view is 6"x9", for observations without adaptive optics. HARMONI's spectroscopic modes have spectral resolving power from 3000 to 17,000 and diagnostic lines, currently expected to including the visible (0.47\,$\mu$m - 0.8\,$\mu$m) and the near-infrared (0.8-2.45$\mu$m).

In addition, two future instruments for ESO's ELT, in the early design phases, will complement existing facitilies for observations of irradiated disks. The MOSAIC instrument (Multi-Object Spectrograph with Adaptive Image Correction) and ANDES (the ArmazoNes high Dispersion Echelle Spectrograph) are anticipated to become operational in the mid-2030s. 
MOSAIC \citep{2024SPIE13096E..15P} will enable the simultaneous observation of over one hundred protoplanetary disks within a field of view of approximately 32 arcmin$^2$. Its 8 mini-IFUs, each featuring a hexagonal field of view of $\approx2.5$" in diameter with spaxels of 150~mas on-sky, will allow highly detailed spectroscopic analyses of those objects. Indeed, MOSAIC is designed to operate across a broad wavelength range, spanning from visible to near-infrared (0.45–1.8 $\mu$m). This wavelength range allows access to tracers morpho-kinematical properties like jets (e.g. FeII), disk (e.g. [CI]) as well as UV irradiation (e.g. OI). 
With options for low ($R \approx 4,000$) and high-resolution ($R \approx 18,000$) spectroscopy, those observations will give access to kinematic properties down to $\lesssim20$~km/s. MOSAIC will thus allow for quick follow-up observations of JWST and VLT/MUSE observed protoplanetary disks and with the sensitivity of such a large telescope, access to fainter spatial properties at the same time. 
ANDES \citep{2024SPIE13096E..13M} is an ultra-stable, high resolving power spectrograph. The instruments is fibre-fed spectrograph and consists of four spectrograph modules that are optimised to different wavelength ranges. Similar to MOSAIC and HARMONI, ANDES will also cover a wide wavelength range (0.4-1.8 $\mu$m, with a goal of extending this to 0.35-2.4 $\mu$m). With R$\sim$ 100, 000, ANDES will add the extra dimension of being able to do detailed kinematics of disks in the key diagnostic lines and, thanks to the inclusion of a small integral field unit coupled to the 1.0-1.8 $\mu$m and fed by an adaptive optics system will add spatial information down to the telescope diffration limit.

\subsubsection{SKA and ngVLA}
The SKA and ngVLA will extend our wavelength coverage at high resolution and sensitivity from the sub-mm regime offered by ALMA to the cm (ngVLA, SKA) and metre (SKA) regimes. 

The SKA \citep{2009IEEEP..97.1482D} will provide huge legacy datasets covering the sky at  wavelengths from $\sim2\,$cm to $6$\,m (50\,MHz--15.4\,GHz).
It will enable the detection of cm sized grains and pebbles in the continuum for the study of solids in disks \citep[e.g.][]{2020MNRAS.498.5116I}. 
These frequency ranges are also well-suited for the detection of radio recombination lines (RRLs), the rotational emission of heavy polyatomic molecules (as the rotational constant is inversely proportional to molecular mass), and the low-frequency hyperfine line emission of key hydride molecules such as OH, CH, and NH$_3$. The high sensitivity and sub-arcsecond angular resolution of these interferometers may enable the detection of complex organic molecules of prebiotic interest that are difficult to detect at higher frequencies in protoplanetary disks \citep[e.g.,][]{2022FrASS...943766J}. The detection of neutral atomic hydrogen (at 1420 MHz), along with the detection of carbon radio recombination lines (RRLs) in proplyds, will allow us to probe the characteristics of the critical H to H$_2$ transition in the PDR of photoevaporating disks. In addition, systematic observations of hydrogen and helium RRLs will reveal the kinematics of their more extended ionization fronts, providing details on how these cocoons interact with their interstellar environment. In general, RRLs are optically thin and can be used to infer electron densities and temperatures (hence, the gas thermal pressure) at the ionization and dissociation fronts,  which are generally largely unknown.

The ngVLA \citep{2017arXiv171109960B, 2018ASPC..517...15S} will bridge the wavelength gap between the SKA and ALMA at high resolution and sensitivity, from 0.26-25\,cm. It will trace free-free emission, and also be an effective tracer of hydrogen radio recombination lines \citep[e.g.][]{2012ApJ...751L..42P}. Combining data from SKA, ALMA and ngVLA to utilise both the high spatial resolution and calculate spectral indices we will be able to clearly distinguish and interpret photoionised winds and other sources of excitation such as jets \citep[][]{2016ApJ...829....1M}.

\subsubsection{ALMA 2030 Roadmap}

 The ALMA Wideband Sensitivity Upgrade \citep[WSU;][]{2020arXiv200111076C} will provide a transformative expansion of the instantaneous correlated bandwidth, reaching at least 16 GHz per polarization across all receivers, along with substantial improvements in receiver and digital path efficiency in the 2030s. The WSU will significantly enhance  sensitivity (essential for studying protoplanetary disks in more distant clusters) and spectral scanning efficiency across all frequencies, with improvements of at least a factor of ten for high-velocity resolution (0.1 km\,s$^{-1}$) scans. This upgrade will make it possible to perform broad-band, unbiased line surveys of proplyds—currently unfeasible within reasonable integration times—using far fewer frequency setups. This will enable multi-transition line detections (yielding accurate molecular excitation and column density measurements) and allow for serendipitous line detections of unexpected chemical species. 
Future improvements into the 2040s may include multibeam and/or on-the-fly  interferometric mapping \citep[e.g.,][]{2010A&A...517A..12P}, which would drastically enhance spatial mapping efficiency and facilitate more unbiased, large-scale surveys of entire star-forming regions (i.e., simultaneously detecting the disk populations and their cloud environment).

\subsubsection{GaiaNIR}
\label{sec:GaiaNIR}
{\it Gaia} has proved extremely valuable for understanding the structure, dynamics and evolution of stellar clusters \citep[e.g.][]{2016A&A...595A...1G, 2018A&A...616A...1G, 2019ApJ...870...32K, 2021ApJ...912..162P, 2024MNRAS.533..705W} and has the potential to do so much more when {\it Gaia} data is fully combined with data at other wavelengths. However, {\it Gaia}'s fundamental limitation is that since it operates in the optical it is significantly limited in regions of higher extinction. {\it GaiaNIR} is the successor to {\it Gaia}, and by operating in the infrared (1--2.5 $\mu$m) would provide astrometry for sources at higher extinctions \citep{2021ExA....51..783H}, opening up the study of the dynamics of both embedded (and therefore younger) clusters and more massive (and typically more distant and more extinguished) clusters, both of which are environments where understanding external photoevaporation is of interest. In clusters that have been well-studied by {\it Gaia} \citep[e.g.,][]{2019ApJ...870...32K,2024MNRAS.533..705W}, {\it GaiaNIR} will extend our reach to fainter, cooler and less-massive stars, which have longer disk lifetimes and where the impact of external photoevaporation can be more pronounced. Finally, {\it GaiaNIR} opens up the possibility of studying the dynamics of stellar clusters in the unique environment of the Galactic centre \citep{2018IAUS..330...67H}. 

{\it GaiaNIR} will achieve a limiting magnitude of $H = 20$ and $K = 20$, detecting an anticipated 10--12 billion stars, 5--6 times that of {\it Gaia} \citep{2021ExA....51..783H}. In addition to detecting more stars, {\it GaiaNIR} observations, anticipated for $\sim$2050, will provide a baseline of over 30 years when combined with {\it Gaia} astrometry. This huge increase in baseline for the vast majority of {\it Gaia} detections will lead to an improvement in proper motion precision of a factor 15 compared to {\it Gaia} alone.

\subsubsection{PRIMA}

PRIMA is a new cooled 1.8\,m FIR space telescope Probe class concept selected for phase A development and providing several orders of magnitude higher line sensitivity than SOFIA or Herschel
\citep[][]{2024AAS...24345002G}. There is currently an absence of FIR facilities. The FIRSS spectrometer \mbox{($\sim$24--235\,$\mu$m)} on board PRIMA will have unique access to HD, a very good proxy of the disk gas mass
\citep{2013Natur.493..644B} and of warmish water vapour \citep[e.g.][]{2021A&A...648A..24V}.
PRIMA will also access the rotational emission of key PDR tracers and dominant and
gas coolants (OH, CH$^+$, CH, high-$J$ CO, [OI]\,63,145$\mu$m, [CII]\,158\,$\mu$m...) as well as the warm dust SED.
PRIMA will provide spatially and spectrally unresolved lines which is not ideal for the study of proplyds, but there is no other way to access those key tracers from the ground and so they would remain valuable.

\section{Summary and Conclusions}
\label{sec:summary}
Motivated by a series of meetings on external photoevaporation of protoplanetary disks we collate thoughts on the past, present and future of observations of externally irradiated disks. We provide a brief review of observations, highlighting the recent flurry of interest and advances in this topic, helped along by recent facilies such as ALMA, JWST and VLT/MUSE. 

We provide an overview of some key outstanding problems in the study of external photoevaporation, which all fall under the key driving questions of how mass is lost in external photoevaporation, what fraction of disks are subject to external photoevaporation at given points in their lifetimes, how does external UV modifies their chemical composition, and whether it has any impact on the planets that form. 

We finish with some suggestions for future observing programs with existing facilities, and a brief look at the potential value of upcoming facilities. Overall there is a lot of work still to be done, but also tremendous potential for progress with existing and upcoming facilities.


\section*{Acknowledgments}
{We thank the anonymous reviewer for their positive review of this manuscript.}

TJH acknowledges funding from a Royal Society Dorothy Hodgkin Fellowship, which funds TP, SP and RM, UKRI guaranteed funding for a Horizon Europe ERC consolidator grant (EP/Y024710/1), which funds LK, SG, LQ and the UKRI/STFC grant ST/X000931/1, which funds GC.   

JRG thanks the Spanish MCINN for funding support under grant PID2023-146667NB-I00.

WJH acknowledges financial support provided by
\foreignlanguage{spanish}{%
  Dirección General de Asuntos del Personal Académico,
  Universidad Nacional Autónoma de México},
through grant
``\foreignlanguage{spanish}{%
  Programa de Apoyo a Proyectos de Investigación
  e Inovación Tecnológica IN109823}''.
  
MCRT acknowledges support by the German Aerospace Center (DLR) and the Federal Ministry for Economic Affairs and Energy (BMWi) through program 50OR2314 ``Physics and Chemistry of Planet-forming disks in extreme environments". 

AJW has received funding from the European Union’s Horizon 2020 research and innovation programme under the Marie Skłodowska-Curie grant agreement No 101104656. 

NPB acknowledges support from NSF grant no. AST-2205698, NASA grant no. JWST-GO-05460, and NASA grant no. JWST-GO-03271. 

RDB acknowledges support from the the Virginia Initiative on Cosmic Origins (VICO) and NSF grant no. AST-2206437. 

AB and JF acknowledge support from the Swedish National Space Agency (2022-00154).

MA, JWE and RJP acknowledge funding from the Royal Society in the form of a Dorothy Hodgkin fellowship and associated enhancement award to RJP.

ADS acknowledges support from the ERC grant 101019751 MOLDISK.

C.W.~acknowledges financial support from the Science and Technology Facilities Council and UK Research and Innovation (grant numbers ST/X001016/1 and MR/T040726/1).

J.~K.~D.-B. acknowledges support from the Science and Technology Facilities Council via a doctoral training grant (grant number ST/Y509711/1).

GB is funded by the European Research Council (ERC) under the European Union’s Horizon 2020 research and innovation programme (Grant agreement No. 853022, PEVAP). 

MLA, KM, CFM are funded by the European Union (ERC, WANDA, 101039452). Views and opinions expressed are however those of the author(s) only and do not necessarily reflect those of the European Union or the European Research Council Executive Agency. Neither the European Union nor the granting authority can be held responsible for them.

JSK acknowledges NASA’s Nexus for Exoplanet System Science (NExSS) research coordination network sponsored by NASA’s Science Mission Directorate and project “Alien Earths” funded under Agreement No. 80NSSC21K0593.



\bibliographystyle{mnras}
\bibliography{references} 

\section*{List of institutions}
$^{1}$Astrophysics Research Cluster, School of Mathematical and Physical Sciences, The University of Sheffield, Hicks Building, Hounsfield Road, Sheffield, S3 7RH, UK\\
$^{2}$Dipartimento di Fisica “Aldo Pontremoli”, Universita degli Studi di Milano, via Celoria 16, Milano, 20133, Italy\\
$^{3}$ European Southern Observatory, Karl-Schwarzschild-Strasse 2, D-85748 Garching bei München, Germany\\
$^4$Astrophysics Group, Department of Physics, Imperial College London, Prince Consort Rd, London SW7 2AZ, UK\\
$^5$Space Science Institute, Boulder, CO 80301, USA \\
$^6$Department of Astronomy, University of Virginia, Charlottesville, VA 22904, USA\\
$^7$Institut de Recherche en Astrophysique et Plan\'etologie, Universit\'e de Toulouse, Centre National de la Recherche Scientifique, Centre National d’Etudes Spatiales, 31028, Toulouse, France\\
$^8$Department of Astronomy, Stockholm University, AlbaNova University Centre, 106 91 Stockholm, Sweden\\
$^9$Astronomy Unit, School of Physics and Astronomy, Queen Mary University of London, London E1 4NS, UK\\
$^{10}$School of Physics and Astronomy, University of Leeds, Leeds, United Kingdom, LS2 9JT\\
$^{11}$Centre for Astrophysics Research, University of Hertfordshire, College Lane, Hatfield, AL10 9AB, UK\\
$^{12}$Dipartimento di Fisica e Astronomia “Augusto Righi”, Università di Bologna, via Gobetti 93/2, 40129 Bologna, Italy\\
$^{13}$Instituto de F\'{\i}sica Fundamental (CSIC). Calle Serrano 121-123, 28006, Madrid, Spain.\\
$^{14}$Istituto Nazionale di Astrofisica (INAF) – Osservatorio Astronomico di Palermo, Piazza del Parlamento 1, 90134 Palermo, Italy\\
$^{15}$ Instituto de Radioastronomia y Astrofisica, Universidad Nacional Autonoma de Mexico, Apartado Postal 3-72, 58090 Morelia, Michoacan, Mexico\\
$^{16}$ Department of Physics and Astronomy, Rice University, 6100 Main
Street, MS-108, Houston, 77005, TX, USA\\
$^{17}$Steward Observatory, University of Arizona, 933 N. Cherry Ave, Tucson, AZ 85721-0065, USA\\
$^{18}$LERMA, Observatoire de Paris, PSL Research University, CNRS, Sorbonne Universités, Paris-Meudon, France.\\
$^{19}$Mullard Space Science Laboratory, University College London, Holmbury St Mary, Dorking, Surrey RH5 6NT, UK.\\
$^{20}$Institut d’Astrophysique Spatiale, Université Paris-Saclay, CNRS, Bâtiment 121, 91405 Orsay Cedex, France\\
$^{21}$Department of Physics and Astronomy, Embry-Riddle Aeronautical University, 3700 Willow Creek
Rd., Prescott, AZ 86301, USA\\
$^{22}$Max-Planck Institut f\"ur Astronomie (MPIA), K\"onigstuhl 17, 69117 Heidelberg, Germany\\
$^{23}$Leiden Observatory, Leiden University, PO Box 9513, 2300 RA Leiden, The Netherlands \\
$^{24}$Space Research Institute, Austrian Academy of Sciences, Schmiedlstr. 6, A-8042, Graz, Austria\\
$^{25}$Instituto de Astrofísica e Ciências do Espaco, Universidade de Lisboa, OAL, Tapada da Ajuda, P-1349-018 Lisboa, Portugal \\
$^{26}$Astrophysics Group, Keele University, Keele, Staffordshire ST5 5BG, United Kingdom \\
$^{27}$ AURA for the European Space Agency, ESA Office, STScI, 3700 San Martin Drive, Baltimore, MD 21218, USA






\label{lastpage}
\end{document}